\documentclass[preprint]{aastex}

\slugcomment{Astronomical Journal, March 2005}
\shortauthors{Fitzpatrick \& Massa}
\shorttitle{Calibration of Synthetic Photometry}

\begin{document}

\newcommand{\atlas}{{\it ATLAS9}}
\newcommand{\synspec}{{\it SYNSPEC}}
\newcommand{\hst}{{\it HST}}
\newcommand{\iue}{{\it IUE}}
\newcommand{\oao}{{\it OAO-2}}
\newcommand{\td}{{\it TD-1}}
\newcommand{\tmass}{{\it 2MASS}}
\newcommand{\hip}{{\it Hipparcos}}
\newcommand{\vsini}{$v \sin i$}
\newcommand{\kms}{km\,s$^{-1}$}
\newcommand{\teff}{{$T_{\mathrm eff}$}}
\newcommand{\logg}{{$\log g$}}
\newcommand{\vturb}{$v_{\mathrm turb}$}
\newcommand{\ebv}{$E$(\bv)}
\newcommand{\grav}{$\log\,g$}
\newcommand{\msun}{${\rm M}_\sun$}

\title{Determining the Physical Properties of the B Stars II. Calibration 
of Synthetic Photometry}

\author{E.L.~Fitzpatrick\altaffilmark{1}, D.~Massa\altaffilmark{2}}
\altaffiltext{1}{Department of Astronomy \& Astrophysics, Villanova 
University, Villanova, PA 19085, USA; fitz@astronomy.villanmova.edu}
\altaffiltext{2}{SGT, Inc., NASA/GSFC, Mailstop 681.0, Greenbelt, 
MD 20771; massa@derckmassa.net}

\begin{abstract}
We present a new calibration of optical ($UBV$, Str\"{o}mgren
$uvby\beta$, and Geneva) and near IR (Johnson $RIJHK$ and \tmass)
photometry for B and early A stars derived from Kurucz (1991) \atlas\/
model atmospheres.  Our sample of stars consists of 45 normal, nearby B
and early A stars which have high quality, low resolution \iue\/ spectra
and accurate \hip\/ parallaxes.  The calibration is unique because it
relies {\em only} on the UV spectral energy distributions, the absolute
flux calibration of the $V$ filter and the \hip\/ distances to determine
the appropriate model atmospheres for the program stars.  These models are
then used to calibrate the synthetic photometry.  We compare our results
with previous, well accepted results and provide a thorough discussion of
the random errors and systematic effects affecting the calibration.  In
particular, we demonstrate the influence of \vsini\/ on surface gravities
derived from fitting model atmospheres.  Finally, we discuss some of our
intended applications of this new calibration.
\end{abstract}
\keywords{  }

\section{Introduction}

We have begun a detailed study of Galactic B and early-A main sequence
stars, whose goals are to both derive detailed measurements of the
basic physical properties of these stars (along with ancillary
information such as distance and interstellar extinction
characteristics) and to critically test the current atmosphere and
interior models which provide the transformation between observable
quantities and the stellar properties, as well as yielding insight into
the structure and evolution of these stars.
 
The B and early-A main sequence stars play a central role in many
stellar astrophysics studies. Simply put, they are the most massive and
luminous objects whose atmospheres are well-represented by the
simplifying assumptions of LTE physics, hydrostatic equilibrium, and
plane-parallel geometry. Thus, precise determinations of their
individual properties should be possible through the application of
currently available and well-developed astrophysical tools.  Because
the evolutionary histories of these objects are generally simple (e.g.,
unaffected by significant mass loss) the interpretation of their
properties --- particularly their surface compositions --- is
straightforward. The potential applications of such results are
manyfold, including tests of stellar structure and evolution
predictions, distance determinations, galactic chemical composition
studies, and the investigation of the physical processes responsible
for chemical peculiarities (notably among the late-B and A types).
Moreover, since the analyses of more massive and/or more evolved
objects are much more physically and computationally challenging
(requiring the inclusion of NLTE physics, dynamical effects, and
evolutionary modifications of their surface abundances), the results
for main sequence B and A stars associated with more massive stars ---
as in a cluster --- provide the essential benchmarks for interpreting
the results for the more luminous objects.
 
As initial steps in this program, we have demonstrated how
low-resolution UV and optical spectral energy distribution (SED)
observations, when combined with the predictions of stellar atmosphere
models, can provide excellent estimates of the basic properties of A or
B stars (see Fitzpatrick \& Massa 1999, hereafter FM). Specifically, we
showed how the SED determines the effective temperature, \teff; surface
gravity, \logg; the abundance of Fe-group elements, [Fe/H]; and the
magnitude of the microturbulent velocity field in the atmosphere,
\vturb. This methodology was then applied to a number of eclipsing
binary systems in both the Milky Way and Large Magellanic Cloud (Guinan
et~al. 1998, 2000; Ribas et al.  2000, 2002; and Fitzpatrick et~al.
2002, 2003).  These results verified that the \atlas\/ (Kurucz 1991)
model atmospheres provide excellent representations of the UV and
optical SEDs of B and A stars and yield reliable estimates of the
stellar properties.  Recently, Niemczura (2003) applied the FM
methodology to slowly pulsating B stars, deriving important results on
the lack of metallicity-dependence of the pulsation phenomenon.

For a few stars, spectrophotometric observations covering the entire
UV-through-optical spectral domain (where B and A stars emit from
70-to-100\% of their energy) are available from spectrometers aboard
the {\it Hubble Space Telescope} (\hst).  These high-precision, long
wavelength baseline, and absolutely calibrated data (see Bohlin 1986),
make the comparison between observations and model predictions
straightforward.  Since both are expressed in flux units, all one has
to do is match the spectral resolution of the observed SEDs with the
theoretical models.  However, to expand our study to encompass large
numbers of stars, we must tap other observational resources. The
primary stellar SED datasets currently available are low-resolution
{\em International Ultraviolet Explorer (\iue)} spectrophotometry
(covering the range 1165-3000 \AA), optical {\it UBV}, {\it
uvby}$\beta$, and Geneva photometry, and near-IR {\it JHK} photometry
from the \tmass\/ project.  To allow these data to be compared with
model atmosphere predictions, the \iue\/ observations must be
absolutely calibrated and the synthetic photometric indices derived
from the models must be transformed to the observational systems.  The
first of these tasks was completed by Massa \& Fitzpatrick (2000), and
this paper addresses the second.

Several calibrations of synthetic optical photometry already exist.
Relyea \& Kurucz (1978) relied on a single star with known physical
properties to calibrate their synthetic $uvby$ photometry, while Balona
(1984) used the Code et al. (1976) sample of stars with known angular
diameters and temperatures determined from UV SEDs to calibrate his
model $uvby\beta$ photometry.  Moon \& Dworetsky (1985) used these same
stars, supplemented by other stars whose physical properties were known
from either model atmosphere analyses or eclipsing binary observations.  
In this way, they extended the range of the calibration and refined its 
accuracy. Later, Napiwotzki et al. (1993, hereafter NSW) employed the Code 
et~al.\ sample, with the \oao\/ fluxes recalibrated by Beeckmans (1977), 
together with A and F stars whose \teff\ and \logg\ values were derived 
from model atmosphere fits to \td\/ observations by Malagnini et al. 
(1982) and lightly reddened B5 -- A0 stars whose physical properties were 
determined from fits to a combination of \td\/ data and optical 
spectrophotometry by Malagnini et al. (1983).  Geneva photometry has been 
previously calibrated by K\"{u}nzli, et al. (1997), among others, and so 
has the Johnson $RIJHK$ photometric system (e.g., Bessell et al.\ 1998) and 
\tmass\/ (Cohen et al. 2003). 

While the goal of this paper is to produce yet another calibration of
synthetic optical and near-IR photometry from the Kurucz \atlas\/ models,
our approach is unique.  Consequently, it provides a valuable
independent verification of previous calibrations and also presents its
own distinct advantages.  Unlike other calibrations, we determine the
interstellar extinction and {\em all} of the physical properties of
each program star (\teff, \logg, [m/H], and $v_{turb}$) by fitting the
details of its \iue\/ UV SED (corrected and placed on the \hst\/ FOS
system using the Massa \& Fitzpatrick 2000, hereafter MF, results) and
its $V$ magnitude to an \atlas\/ model.  Furthermore, the properties of
each star are constrained by a combination of the observed \hip\/
distance to the star and stellar interior models.  The best fitting
\atlas\/ model for each star is then used to calculate its synthetic
photometry.  Finally, the synthetic model photometry of all the program
stars is fit to their observed photometry in order to derive the
transformations.  The advantages of this approach are that it utilizes
more spectral information than previous calibrations (allowing an
unambiguous determination of the physical properties for each star),
and that it incorporates independent, non-spectroscopic ancillary data
directly into the calibration process.  In addition, this calibration
provides the most internally consistent transformation when UV SEDs and
optical and near-IR photometry are fit simultaneously.

In the sections below, we begin with a general discussion of our
calibration strategy (\S~2), followed by a description of the main
ingredients in the analysis, i.e., the sample of stars and the adopted
grid of stellar atmosphere models (\S~3).  We then, in \S~4, present the
results of the model-fitting procedure and assess the internal sources
of error. We describe how the synthetic photometry is calibrated in
\S~5 and present the final catalog of calibrated synthetic photometry in
\S~6.  In \S~7, we compare the calibrated data to other, currently
available calibrations, assess possible sources of systematic errors,
and describe how to apply the calibrated photometry.  Finally, in \S~8,
we summarize our results and highlight several potential applications.

\section{Overview of the Photometric Calibration}

Our calibration of synthetic photometry involves three steps:  First,
we find the reddened and distanced-attenuated stellar atmosphere models
which best match the observed SEDs of our program stars, as described
below.  Second, we perform synthetic optical and near-IR photometry on
the models for each program star to obtain a set of uncalibrated
synthetic photometric magnitudes and indices for each star.  Third, we
compare the synthetic photometry with the observed photometry, to
determine the transformations between the observed and synthetic
systems.  This latter step is essentially identical to that performed
in standard photometric reductions, where instrumental values are
transformed to a calibrated standard system.

We model the shapes of the stellar SEDs as follows.  Let $f_{\lambda}$
be the SED of a star as observed from Earth. It depends on the surface
flux of the star and on the attenuating effects of distance and
interstellar extinction and can be expressed as
\begin{eqnarray}
f_{\lambda} &=& F_{\lambda} \times \left(\frac{R}{d} \right)^2 \times 
10^{-0.4 E(B-V) [k(\lambda-V) + R(V)]}
\label{basic}
\end{eqnarray}
where $F_{\lambda}$ is the surface flux, $R$ is the stellar radius, and
$d$ is the distance.  The last term carries the extinction information,
including  $E(B-V)$, the normalized extinction curve $k(\lambda-V)
\equiv E(\lambda-V)/E(B-V)$, and the ratio of selective-to-total
extinction in the $V$ band $R(V) \equiv A_V/E(B-V)$.  We use the MPFIT
procedure by Craig Markwardt to determine the optimal values of all the
parameters which contribute to the righthand side of equation
\ref{basic}, and thus to achieve the best fit to the observed fluxes
$f_{\lambda}$ (see Markwardt 2003).
 
The observed SEDs, $f_{\lambda}$, consist of \iue\/ UV
spectrophotometry (1165 -- 3000 \AA) and the optical $V$ magnitudes,
calibrated according to FM.  As in FM, a weight is assigned to each
\iue\/ wavelength bin.  This is a combination of the statistical errors
in the data for that bin summed quadratically with an additional,
systematic, uncertainty of 2.5\% which accounts for residual
low-frequency uncertainties in the data.  We assign zero weight to the
wavelength region 1195 -- 1235 \AA\/ due to the presence of
interstellar Ly~$\alpha$ absorption. The $V$ magnitude is assumed to be
exact (however, see \S \ref{secERR}).  We represent the stellar surface
fluxes $F_{\lambda}$ by \atlas\/ model atmosphere fluxes, which are
functions of four parameters:  \teff, \logg, [m/H], and \vturb.
Because all of our program stars are lightly- or un-reddened we can
utilize a predetermined form for the shape of the interstellar
extinction curve and introduce very little error.  All but the most
deviant extinction curves differ by only a few magnitudes from the
Galactic average curve.  While this can result in huge effects for
stars with significant extinction, for stars with color excesses of a
few hundredths or less (as is the case here), the effect is minimal.
Consequently, we adopt the $R(V) = 3.1$ extinction curve from
Fitzpatrick (1999) to represent the shape of the extinction law.  With
these assumptions, we see that six parameters --- \teff, \logg, [m/H],
\vturb, the attenuation factor $(R/d)^2$, and \ebv ---  must be
specified to completely define the SED of a lightly reddened star.

The data we utilize in the fitting process (\iue\/ spectrophotometry
and the $V$ magnitude) strongly constrain three of the four parameters
needed to specify the best-fitting \atlas\/ models, namely, \teff\/,
[m/H], and  \vturb, but are relatively insensitive to the fourth
parameter, \logg.  Consequently we require additional information to
determine \logg.  Normally, this could be achieved by comparing
high-resolution spectroscopic observations of a Balmer line profile
with Stark broadened model profiles.  Unfortunately, we do not have
such measurements at our disposal, and must look elsewhere.  Our
solution is to employ a combination of \hip\/-based distances and
stellar structure models, in conjunction with the SED data, to provide
a very strong constraint on $\log g$ without appealing to hydrogen line
data.

\begin{figure}[ht]
\figurenum{1}
\epsscale{0.7}
\plotone{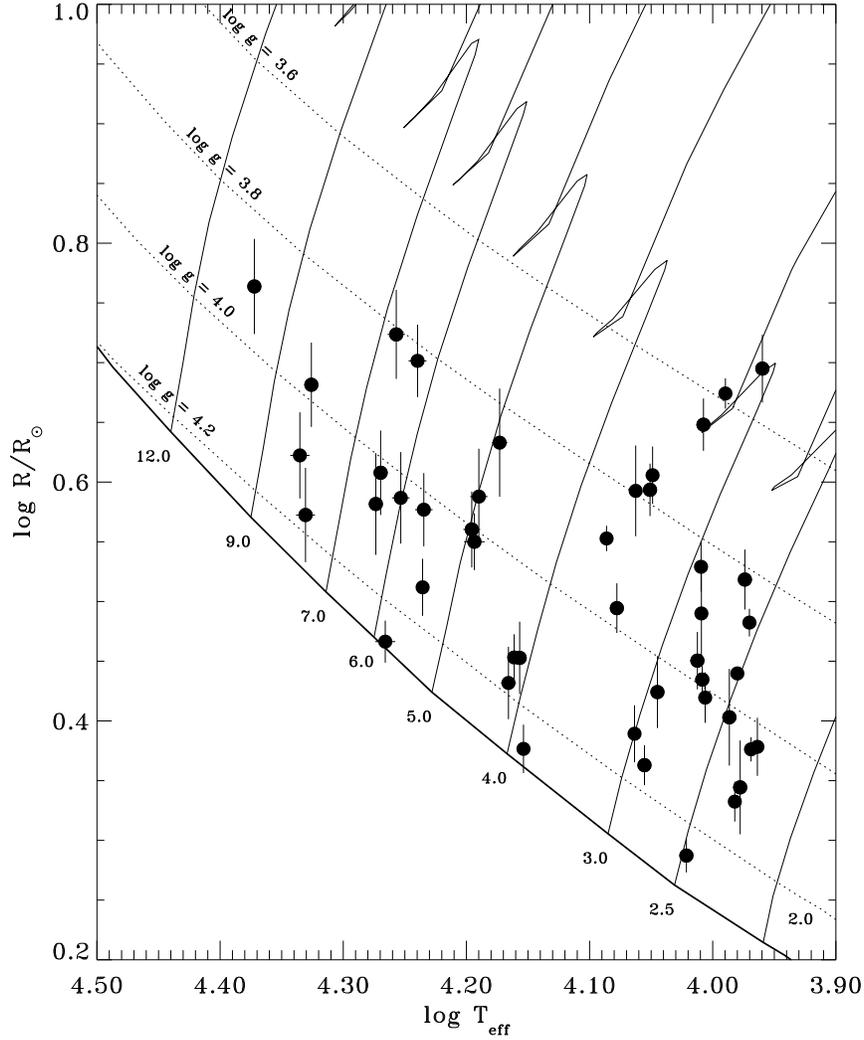}
\caption{Theoretical HR Diagram, expressed in terms of $\log R/R_\odot$
vs $\log T_{eff}$, based on the stellar evolution models of Bressan et
al.  1993.  The thick solid curve shows the location of the Zero Age
Main Sequence (ZAMS) and the thin solid curves show the evolution
tracks for the initial stellar masses listed along the ZAMS, in units
of $M/M_\odot$.  Lines of constant Newtonian surface gravity (i.e.,
$GM/R^2$) are shown for several representative and relevant values.
The filled circles show the positions of our program stars in the HRD.
Note that, for most stars, the error bar for $T_{eff}$ lies within the
symbol. The derivation of the stellar properties is discussed in \S\S\/
2 and 4.
\label{figHRD}}
\end{figure}

Our approach is to adopt the distances $d$ implied by \hip\/ parallax
measurements, which then allows us to determine the stellar radii $R$
directly from the SED fitting procedure, rather than the ratio $R/d$
(see Eq. \ref{basic}). For main sequence stars, stellar structure
models show that the surface gravity is uniquely identified by a star's
\teff\/ and $R$.  This is demonstrated in Figure \ref{figHRD} using the
Padova grid of structure models (Bressan et~al.  1993) in the mass
range of interest.  Thus, by incorporating the distance constraint, we
can fit the program star SEDs with five parameters (\teff, [m/H],
\vturb, $R$, and \ebv) and then determine the surface gravity from
stellar structure models {\it via}\/ $\log g = \log g(T_{eff},R)$.
However, the surface gravity returned by the structure models is the
``Newtonian'' gravity,
\begin{eqnarray}
g(Newton) \equiv G\frac{M}{R^2}\/,
\label{gnewton}
\end{eqnarray}
which is not appropriate for describing the emergent flux of even
moderately rotating stars, since the centrifugal force reduces the
atmospheric pressure below that implied by $g(Newton)$.  To compensate
for this effect we make a first-order correction to $g(Newton)$ to
derive the ``spectroscopic'' gravity,
\begin{eqnarray}
g(Spec) \equiv g(Newton) - \frac{(v \sin i)^2}{R}\/.
\label{gspec}
\end{eqnarray} 
This corrected value of the gravity is used to select the appropriate
\atlas\/ model and the whole process is iterated to achieve consistency.
As a result of this procedure, we are able to derive well-defined
estimates of all five fitting parameters, plus $\log g(Spec)$, from
just the UV SED, $V$, $v \sin i$, and the \hip\/ distances.

To our knowledge, Herrero et al. (1992) were the first to formally
correct spectroscopically-derived surface gravities for the effects of
rotation --- although such effects on gravity-sensitive spectral
features have been recognized for some time (see, e.g., Gray \&
Garrison 1987). The form of the correction in  Eq. \ref{gspec} is based
on the detailed discussion of Repolust et al.\ (2004).  We emphasize
that this a first-order correction to a complex effect and that more
sophisticated approaches exist for interpreting the spectra of rapidly
rotating stars (e.g., Collins \& Truax 1995, Towsend et al.\ 2004).
However, these results are highly model-dependent, are more appropriate
for extremely rapidly rotating stars (which are not included our
sample), require detailed line profiles for analysis (which we do not
possess), and have never been fully verified.  In contrast, the
simplified adjustments we employ are easily applied and will be shown
in \S~7.2 to be entirely consistent with observations.

\section{Data}

\subsection{The Observations}

For this study, we restrict our attention to normal, non-supergiant,
non-emission line, lightly-reddened ($E(B-V) \leq 0.03$) A and B stars
with high quality low-resolution \iue\/ spectra (Boggess et~al.\ 1978)
for both the short (SWP) and long (either LWP or LWR) wavelength
ranges, high quality optical photometry, and \hip-based distances with
errors $\leq 10$\% (Perryman et~al. 1997).  These criteria resulted in
the sample of 45 stars listed in Table~\ref{tabSTARS}.

We use NEWSIPS \iue\/ data (Nichols \& Linsky 1996) obtained from the
MAST archive at STScI.  These data were corrected for residual
systematic errors and placed onto the \hst/FOS flux scale of Bohlin (1996)
using the corrections and algorithms described by Massa \& Fitzpatrick
(2000; hereafter MF).  Multiple spectra from each wavelength range (SWP
or LWR and LWP) were combined using the NEWSIPS error arrays as
weights.  Small aperture data were scaled to the large aperture data
and both trailed and point source data were included.  Short and long
wavelength data were joined at 1978~\AA\ to form a complete spectrum
covering the wavelength range $1150 \leq \lambda \leq 3000$~\AA.  Data
longward of 3000~\AA\ were ignored because they are typically of low
quality and subject to residual systematic effects.  The \iue\/ data
were resampled to match the wavelength binning of the \atlas\/ model
atmosphere calculations in the wavelength regions of interest.

Note that the MF corrections to the \iue\/ NEWSIPS data are essential
if precise absolute fluxes are required --- as in the present study ---
since the systematic errors in these data are wavelength-dependent and
approach 10\% in some spectral regions.  Although we have not performed
a detailed examination of the European Space Agency's version of \iue\/
Final Archive (i.e., the INES database), spot examination of a number
of INES spectra has shown that similar systematic problems affect these
data.  As with NEWSIPS, INES fluxes should be considered with caution,
as the absolute calibration appears to be systematically in error and
residual thermal and temporal affects may exist in the data.

Table~\ref{tabPHOT} lists the Johnson {\it UBVRIJHK} and Str\"{o}mgren
{\it uvby}$\beta$ photometry for the program stars and
Table~\ref{tabGEN} lists the available Geneva color indices.  All data
in Tables~\ref{tabPHOT}~and~\ref{tabGEN} are mean values acquired via
the Mermilliod et~al. (1997) archive.  Table~\ref{tab2MASS} lists the
\tmass\/ {\it JHK} data along with their associated errors.  These data
were obtained from the \tmass\/ All-Sky Point Source Catalog at the
NASA/IPAC Infrared Science Archive.

\subsection{The Models}

The model data consist of stellar surface fluxes $F_{\lambda}$ produced
by the Kurucz (1991) \atlas\/ model atmosphere code, in units of
erg~cm$^{-2}$~sec$^{-1}$~\AA$^{-1}$.  Since the resolution of the
\iue\/ (4.5 -- 8 \AA) is only slightly smaller than the size of
the \atlas\/ wavelength bins in the UV region (10~\AA), the
characteristics of binned \iue\/ data do not exactly match those of
\atlas, where the latter may be thought of as resulting from the
binning of very high resolution data.  In essence, adjacent \iue\/ bins
are not independent of each other, while adjacent \atlas\/ wavelength
points are independent.  To better match the \iue\/ and \atlas\/ data,
we smooth the \atlas\/ model fluxes in the \iue\/ wavelength region by
a Gaussian function with a FWHM of 6~\AA.  This slight smoothing
simulates the effects of binning lower resolution data.  FM have shown
that these provide excellent fits to UV data.
 
\section{Model Fitting}

In this section, we present the results of the our fitting procedure and 
assess their observational errors. 
 
\begin{figure}[ht]
\figurenum{2}
\epsscale{0.75}
\plotone{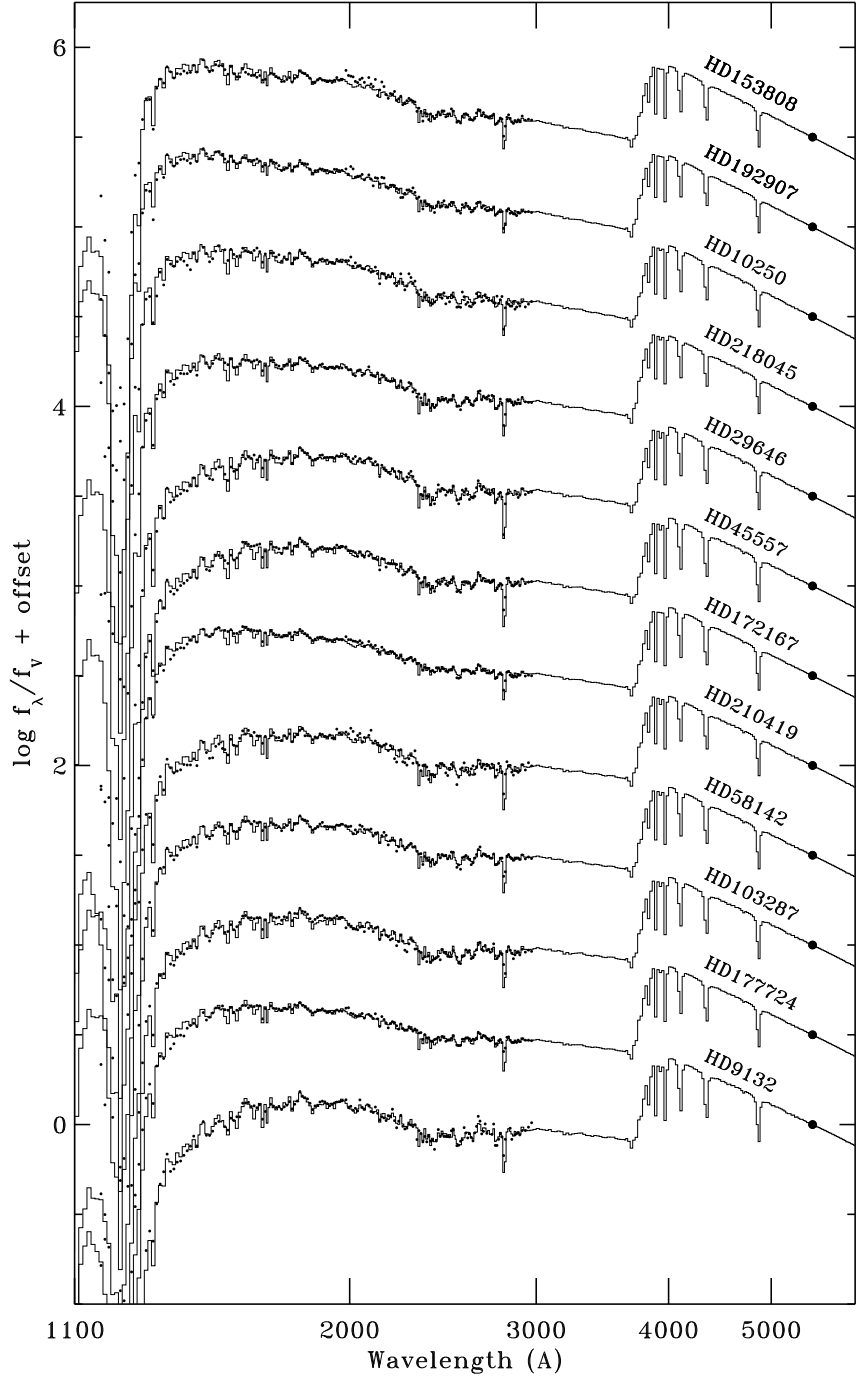}
\caption{\scriptsize{Model fits to the observed fluxes of the program stars.  The
solid histogram-style curves are the best fitting models (reddened to
match the observations).  The points are the observed \iue\/ data and
$V$ magnitudes.  The x-axis is a logarithmic wavelength scale.  The
y-axis is $\log f_{\lambda}/f_V$, with each star offset for display
purposes.  Each tick mark on the ordinate is 0.5 dex.  Each star is
labeled with its HD number.  The fits are arranged so that \teff\/
increases upward in each plot and with each subsequent plot.}
\label{figSEDa}}
\end{figure}

The four panels of Figure~\ref{figSEDa} illustrate the results of the SED 
fitting procedure for the 45 program stars.  The \iue\/ (small circles) 
and $V$-band (large circles) data are shown, along with the best-fitting 
distance-attenuated and reddened \atlas\/ models (histogram style curves).  
The stars are arbitrarily offset vertically for display purposes and are 
arranged in order of increasing \teff, with the coolest at the bottom of 
the first panel and the hottest at the top of the last panel.  As
has been shown already by FM, Guinan et~al. (1998, 2000), Ribas et~al.
(2000, 2002), and Fitzpatrick et~al. (2002, 2003), the ability of the
\atlas\/ models to reproduce the UV SEDs is impressive.

Values of the fit parameters are listed in Table~\ref{tabPARMS}.
Recall, from the discussion in \S~2, that the first five parameters in
the Table (\teff, [m/H], \vturb, $R$, and \ebv) are determined directly
by the fit, while the sixth one ($\log g(Spec)$) is deduced utilizing
the \hip\/ distances, $v \sin i$, and the Bressan et al.\ models.  For
completeness, we also list the value of the Newtonian gravity ($\log
g(Newton)$) in the final column of the Table.  To provide an overview of
the properties of our calibration sample, we plot their temperatures
and radii (plus the associated errors) on the theoretical HR Diagram
shown in Figure \ref{figHRD}.

\subsection{The Uncertainties}\label{secERR}

The errors listed in Table~\ref{tabPARMS} are 1-$\sigma$ and
incorporate a number of potential sources of uncertainty.  The MPFIT
least-squares routine provides estimates of the internal uncertainties
for each of the parameters, based on the statistical errors in the
input data (in this case, the \iue\/ SEDs) and any covariant behavior
of the parameters.  However, additional uncertainty in the results
arises from uncertainty in some of the assumptions in the analysis,
which are not communicated to the fitting routine.  These include
random errors in the $V$ magnitudes (which are assumed to be exact by
the fitting algorithm) and in $v \sin i$; scaling errors in the \iue\/
fluxes which translate to zero point offsets in their logarithms (see
MF); uncertainties in the \hip\/ distances; and errors in the shape of
the assumed extinction curve and its value of $R(V)$.

We incorporated these possible effects in our analysis using a Monte
Carlo approach.  The SED of each star was fit 100 times, with input 
parameters (i.e., $V$, \iue\/ zero points, $d$, $k(\lambda-V)$, and
$R(V)$) drawn randomly each time from parent samples generated by 
assuming Gaussian-like error distributions.  These parent distributions 
were assumed to have the following properties:  
\begin{enumerate}
\item  All $V$ magnitudes have 1-$\sigma$ observational errors of
  $\pm$0.015 mag.
\item  All $v \sin i$ values have 1-$\sigma$ observational errors of $\pm$10\%.
\item  The logarithmic uncertainties in zero points of individual \iue\/ 
  spectra correspond to the scaling errors given in Table~5 of MF.  In 
  cases where multiple spectra were combined, the uncertainties are 
  reduced accordingly.  For example, a single LWR large aperture spectrum 
  has a 1-$\sigma$ zero-point uncertainty of $\pm$3.8\%; when two such 
  spectra are combined, the uncertainty reduces to $\pm$2.7\%.  
\item The 1-$\sigma$ errors in the \hip\/ distances are those listed in 
  Table~\ref{tabSTARS}.
\item  The 1-$\sigma$ uncertainty of $R(V)$ is $\pm$0.4 mag, and the 
  corresponding changes in the shape of the extinction curve are 
  calculated from the prescription given by Fitzpatrick (1999).  This 
  error in $R(V)$ produces a change in the normalized extinction at 
  1550~\AA\ of 0.69~mag, which slightly exceeds the RMS scatter found in 
  randomly selected extinction curves for low density sight lines 
  (see, Table II in Massa 1987).
\end{enumerate}

For each star, the uncertainties listed in Table~\ref{tabPARMS} were
computed by quadratically combining the internal errors returned by
MPFIT with the standard deviation observed for each output parameter
within a set of 100 Monte Carlo SED fits.

Other potential sources of error include the adopted stellar interior
models, which provide the \logg\/ values, and the \atlas\/ stellar
atmosphere models themselves.  We tested the former by running the
analysis utilizing the Geneva grid of stellar interior models (Schaller
et~al. 1992).  This had virtually no effect on our results, since
typical differences between the Geneva-based and Padova-based
parameters are an order of magnitude (or more) smaller than the
parameter uncertainties listed in Table~\ref{tabPARMS}.  The issue of
the precision of the \atlas\/ model is beyond the scope of this paper.
Here we can merely note that these models obviously yield extremely
good fits to the observations (see Figure 1) and, in cases where
substantial ancillary information exists (e.g., for eclipsing binary
systems, see Fitzpatrick et~al. 2003), they appear to yield precise and
accurate values of the stellar properties.

A look at the results in Table~\ref{tabPARMS} shows that the stellar
properties are apparently very well-determined.  Some faith in the
results and the quoted uncertainties can be gained by considering the
case for Vega (HD 172167), for which we find $T_{eff} = 9549\pm41$,
$\log g = 3.96\pm0.01$, and ${\rm [m/H]} = -0.51\pm0.07$.  The error
bars for our parameters are remarkably small, but the results are in
excellent agreement with those from other investigators (see, e.g.,
Castelli \& Kurucz 1994, Qiu et~al. 2001) and consistent to within
the stated errors.   In \S~\ref{secDISCUSSION}, we provide additional 
evidence that our procedure has indeed yielded reasonable values for the 
stellar properties and that these are consistent with determinations from 
previous studies.

\clearpage
\section{The Synthetic Photometry}\label{secSYNPHOT}

This section describes the calculation and calibration of the synthetic 
photometry and presents the final results.  

\subsection{Calculation}

Now that we have reliable SEDs with full wavelength coverage, all that
is needed to compute the synthetic photometric magnitudes and indices
is a set of sensitivity curves for the various filters.
Table~\ref{tabFILT} lists the references for the sensitivity curves we
have adopted. The uncalibrated synthetic photometry was calculated by
convolving the model SED for each star with the appropriate filter
sensitivity curve, scaled to unit area, and converting the result to
magnitudes.  The uncalibrated indices, such as $(B-V)_{syn}$, were
computed by differencing the appropriate magnitudes.  For the Johnson
$B-V$ and $U-B$ indices, we followed the procedure of Buser \& Kurucz
(1978) and computed $(B-V)_{syn}$ with their $B_3$ and $V$ filters and
$(U-B)_{syn}$ with their $U_3$ and $B_2$ filters, using their naming
convention.

Synthetic Str\"{o}mgren $\beta$ photometry requires some additional
discussion.  The $\beta$ index --- which measures the strength of the
H$\beta$ line at 4861 \AA\/ and is very sensitive to surface gravity
--- is obtained by computing the difference between an
intermediate-band magnitude and a narrow-band magnitude, both centered
on H$\beta$.  The width (FWHM) of the intermediate filter is typically
in the range 90--150 \AA\/ while that of the narrow filter is 15--35
\AA.  We experimented with a variety of profiles for these filters,
including those published by Crawford \& Mander (1966) and actual
profiles of filters currently in use at Kitt Peak Observatory.  The
very strong result of these experiments is that our final results are
virtually independent of which filter set we choose for the synthetic
photometry since, as in the case of real photometry, a transformation
relation maps the synthetic photometry onto the standard system.  The
different filter combinations do, of course, lead to different
transformations and we based our choice of filter on the shape of the
relation.  Our ultimate choices were Gaussian-shaped
filters with FWHM values of 90 \AA\/ and 15 \AA, which are reminiscent
of the filters used by Crawford (1958) in the early establishment of the
$\beta$ system and which allow a simple linear transformation to the
standard system. These filters are shown in Figure \ref{figBETA},
compared to a synthetic H$\beta$ profile corresponding to a main
sequence star with \teff = 15000 K. 

Because of the narrowness of the $\beta$ filters, the synthetic
photometry cannot be performed directly on the low-resolution \atlas\/
SEDs, which are binned at 20~\AA\/ intervals near H$\beta$.  Instead,
we first computed high-resolution synthetic spectra for each \atlas\/
model --- sampled at 0.1~\AA\ over a 400~\AA\ interval centered on
H$\beta$ --- utilizing the \synspec\/ spectral synthesis program
(Hubeny \& Lanz 2000).  We then calculated the synthetic photometry
using these spectra and with the filter profiles scaled to peak values
of unity.  The  H$\beta$ line seen in Figure \ref{figBETA} is from this
set of calculations.

\subsection{Calibration}\label{secCAL}

The last step in the calibration of the synthetic photometry is
determining the functional relationship between the synthetic
magnitudes and indices and the observed photometry for the 45 program
stars.  Figures 4 through 7 show these transformation relationships for
the Johnson, Str\"{o}mgren, and Geneva filters. In each panel of the
figures, the solid line has a slope of unity and shows the mean offset
between the observed and synthetic values. In cases where such a simple
offset is deemed insufficient to described the relationship between
observed and synthetic values, the adopted transformation is shown by a
dashed line.  The standard deviations ($\sigma$), or ``scatter,'' of
the observed indices around the adopted transformation lines are indicated
in the lower right of each panel of Figures 4--7.

For $B-V$, $m_1$, $(V-B)_{Geneva}$, $(G-B)_{Geneva}$, $J$, $H$, and
$K$,  a simple offset provides the best transformation relationship, with
little or no improvement from the use of higher order terms or color
terms.  All the other colors and indices required a linear term,
indicated in their respective panels by the dashed lines.

Note that the $\beta$ index is well represented by a linear
transformation, although the dynamic range of the synthetic index is
somewhat larger than for the standard system (i.e., the slope of the
transformation is less than 1.0).  We found that we could steepen the
relation to better match the standard system by, for example, widening
the adopted 15 \AA\/ filter.  However, this always resulted in
significant curvature in the transformation {\em with no improvement in
the scatter of the observed data about the transformation curve} and we
opted for the simpler transformation.  As will be shown below, the
observed scatter is consistent with the known sources of random error.
   
Figure~\ref{fig2MASS} shows the differences between the observed and
synthetic \tmass\/ magnitudes plotted against the observed values.  The
error bars are for the observed photometry, as listed in
Table~\ref{tab2MASS}.  For each filter there are clearly two distinct
groups of data points, one well-determined and one poorly-determined.
This arises because the images of our brightest stars are highly
saturated in \tmass\/ observations and the derived magnitudes are
highly uncertain, with 1-$\sigma$ errors of $\sim$0.2--0.3~mag.  For
the fainter stars, the errors are a factor of 10 smaller.  The
horizontal dashed lines in Figure~\ref{fig2MASS} show the simple mean
offset between the observed and synthetic magnitudes, based only on the
well-determined observations.  No higher order terms are required in
the transformation and it is gratifying that there appears to be no
systematic difference between the results for the brightest stars and
the rest of the \tmass\/ sample.  The standard deviations of the
observed data about the mean transformation lines are indicated in each
panel of Figure~\ref{fig2MASS}, computed only from the data with errors
less than 0.1 mag.
 
The detailed level of agreement between the observed and synthetic
photometric indices, i.e., the ``scatter'' about the mean
transformation lines in Figures 4--\ref{fig2MASS}, provides some
insight into the accuracy of our procedure.  If the calibrations are 
correct, then the scatter of the observed data about the transformation 
relations should result from the simple quadratic sum of the 
observational errors and the random errors in the synthetic photometry.  
Thus, comparing the observational errors, synthetic photometry errors, 
and observed transformation scatter allows us to examine whether the 
error estimates for the stellar properties are reasonable (see 
\S~\ref{secERR}) and whether some additional sources of error are 
present in the procedure.

The data for these comparisons are summarized in Table~\ref{tabSCAT}.
The second column lists the transformation scatter for each index (as
shown in Figures~4--\ref{fig2MASS}).  The third column gives the
expected errors in the synthetic indices.  These arise from the
uncertainties in the SED fitting procedure (e.g., the uncertainty in
the best-fitting values of $T_{eff}$) and were estimated for
each star by performing synthetic photometry on each of the 100 Monte 
Carlo fits described in the previous section.  This resulted in standard 
deviations for each synthetic photometric index in each star.  We then 
computed the RMS mean of these standard deviations for each index, 
averaging over all the stars.  This provides an estimate of the 
characteristic uncertainty of the synthetic photometry for our program 
stars.  Finally, the quadratic difference between the transformation 
scatter and the synthetic photometry error provides an estimate of the 
actual observational errors in the data, in the absence of any other
significant source of random error.  This estimate is given on the
fourth column of Table~\ref{tabSCAT}.

Examination of the results in Table~\ref{tabSCAT} shows that the
observational uncertainties implied by the scatter seen in our
transformation relations are completely consistent with expected
uncertainties, which are listed in the final column of the table.
Among the optical Johnson, Str\"{o}mgren, and Geneva indices, only the
Geneva $U-B$ color shows a significantly larger scatter than expected.
However, we suspect that the ``expected'' value is somewhat optimistic
given the inherent difficulties with atmospheric extinction
corrections for the broadband $U$ filter.  This effect leads to the
larger scatter (both observed and expected) for the Johnson $U-B$ color
and probably affects the Geneva $U-B$ color in a similar way.  We are
not aware of analyses of the expected uncertainties in Johnson $V-R$
and $R-I$ photometry. However, given the great breadth of the filters,
and the wide range in the properties of the filters actually in use
--- both of which compromise the transformability of the photometry ---
the ``predicted'' scatter of 0.02--0.03 mag seems reasonable. The small
number of program stars for which Johnson $JHK$ data are available make
the derived estimates of the observational scatter uncertain, but the
results are clearly consistent with errors of 0.01--0.02 mag, which are
also reasonable.

For the \tmass\/ data, we can compare the predicted observational
errors directly with the errors provided with the data.  These two sets
of results are listed in Table~\ref{tabSCAT}, utilizing only the
``well-determined'' observations, as described above.  For $J_{2M}$ and
$K_{2M}$ the results are consistent, but the calibration scatter for
$H_{2M}$ is significantly larger than can be accounted for by the
quoted observational errors and our expected synthetic photometry
errors.  Since the $H_{2M}$ band straddles the head of the hydrogen
Brackett series it seemed possible that inadequacies in the \atlas\/
SEDs could produce larger random errors than we computed if, for
example, the strengths of the higher order Brackett lines were more or
less temperature-sensitive than predicted by the models.  However, we
find no evidence in our results for such an effect.  The $H_{2M}$
transformation scatter for the cooler stars, where the Brackett lines
are stronger, is actually slightly smaller than for the hotter stars,
where the lines are intrinsically weaker.  We tentatively conclude that
the observational errors for $H_{2M}$, which are already larger than
for $J_{2M}$ and $K_{2M}$, may be underestimated, perhaps due to
larger-than-expected variations in atmospheric transmission in this
band.  Note that the transformation scatter for the Johnson $H$ is also
larger than for the nearby $J$ and $K$ bands, also indicting larger
observational uncertainties.

Recently, Cohen (2003) published a detailed calibration for the
\tmass\/ filters, based on a sample of 33 A0-through-M0 stars.
Although the dynamic range of this sample greatly exceeds ours --- in
terms of both magnitude and color --- the results are quite
compatible.  In the simplest terms, $J$, $H$, and $K$ magnitudes
derived using our calibration can be transformed to Cohen's system by
adding 0.024, 0.007, and 0.023 mag, respectively for the three
filters.  This consistency is a testament to the uniformity of the
\tmass\/ database and the accuracy of the \tmass\/ filter sensitivity
curves.
 
Our overall conclusion from considering the results in
Table~\ref{tabSCAT} and Figures~4--\ref{fig2MASS} is that the synthetic
photometry based on the \atlas\/ models is well-behaved and readily
transformed to the standard photometric systems.  Further, the
consistency between the expected scatter in the calibration
relations and that actually measured indicates that our error analysis
has provided reasonable estimates of the uncertainties in the
parameters describing the models and thus, the stellar properties.  

\section{Final Results}\label{secRESULTS}

The captions to Figures 4--\ref{fig2MASS} contain the
actual transformation relations for all the photometric indices;
however, these are not particularly useful to the reader since they
depend in detail on exactly how we computed the synthetic values.  Of
more use are the actual calibrated synthetic indices themselves.  In
Tables~8,~9,~10,~and~11, we present the calibrated photometry for all
the models in our \atlas\/ grid for the Johnson, Str\"{o}mgren, Geneva,
and $2MASS$ systems, respectively.  As described by FM, this grid
consists of models with \teff\/ between  9000 K and 50000 K and \logg\/
from 5.0 down to the Eddington limit for five different values of the
metal abundance ([m/H] = --1.5, --1.0, --0.5, 0.0, 0.5) and five
different microturbulence velocities (\vturb\/ = 0, 1, 2, 4, 8 \kms).
There are a total of 8847 models in the grid. Each of the tables is
subdivided into 25 parts (e.g., ``8a'', ``8b'', etc.), with each part
containing the photometric indices for one combination of [m/H] and
\vturb.  In the interest of saving space, we show here only a small
portion of Tables 8a, 9a, 10a, and 11a.  The complete versions of the
tables are available in the electronic edition of the Journal.

\section{Discussion \label{secDISCUSSION}}

Our goal has been to derive a calibration of synthetic photometry from
the \atlas\/ model grid which will enable us to infer the physical
properties of a star and its line-of-sight extinction from observations
of its UV continuum and optical photometry.  To determine the
calibration, we constrained the models by demanding that the derived
physical properties be consistent with a set of conditions imposed by a
combination of \hip\/ data and stellar structure models.

Now that we have a calibration of the optical and NIR photometry, three 
issues arise:  
\begin{enumerate}
  \item {\it Are the fits and the derived physical properties reasonable?}
  \item {\it How well can the stellar properties be determined?}
  \item {\it How is the best fit determined when \hip\/ data are 
    unavailable?}
\end{enumerate}
To answer the first question, we compare our derived stellar properties
to previous, accepted values, as assigned to their spectral types or
derived from their $uvby\beta$ photometry.  To answer the second question, 
we must evaluate the influence of both measurement and systematic errors 
on the derived parameters.  The third question amounts applying the 
calibration to the synthetic photometry and then re-deriving the physical 
properties from the UV and optical photometry alone.  This process also 
forces us to address the question of how to assign weights to the UV and 
optical photometry.

\subsection{Comparison to previous results}

It is important to verify our fitting procedure by comparing the
physical properties we derive with those obtained by  previous
investigators.  We begin by examining the relationship between \teff\/
and MK spectral type.   Figure \ref{spty}, displays the derived
temperatures versus the spectral types for the program stars (filled
circles).  There is only one disparate point, and that is for the
hottest star in our sample, $\zeta$~Cen (HD~121263), which is
classified as B2.5~IV by Hiltner, Garrison, \& Schild (1969).  However,
the low resolution UV spectrum of this star is nearly identical to that
of the unreddened B1~V star HD~31726 (too distant to be in the current
sample), and differs significantly from other B2 stars in our sample.
Furthermore, this star has been previously classified as B1~V by
Woolley, Gascoigne, \& de~Vaucouleurs (1954) and its $uvby\beta$
photometry (see below) is also consistent with a B1~V star.  Thus, we
suspect a classification error in this case.  Figure \ref{spty} also
shows a number of spectral type vs.  $T_{eff}$\/ calibrations taken
from the literature (and identified in the figure).  It is clear that
our results are quite consistent with past evaluations.

A commonly used method for calculating the temperatures and surface
gravities of A and B stars is to derive them from Str\"{o}mgren
$uvby\beta$ photometry.  Because the photometry does not provide
adequate information to determine [m/H] or \vturb\/ for B stars, these
are typically either fixed or derived from a fine analysis of the
spectrum.  Napiwotzki (2004) provides an updated version of the widely
used Moon \& Dworetsky (1985) synthetic photometry calibration, which
is also based on \atlas\/ models.  We have used this program to
calculate \teff\/ and \logg\/ from $uvby\beta$ photometry for each
program star.  We plot these photometry-based properties against ours
in Figures \ref{compare}.  It is immediately apparent that the
agreement is quite good, with very small systematic differences or
scatter.  The mean differences (ours minus $uvby\beta$) and their RMS
scatters are:  $\Delta \log T_{\rm eff} = 0.000 \pm 0.009$ and
$\Delta \log g = 0.03 \pm  0.13$, indicating that the small offsets
are well within the scatter.

It should be emphasized that the two \teff\/ and $\log g$
determinations presented in Figures \ref{compare} {\em have no input
data in common.}  The properties we derive are based on absolutely
calibrated UV spectrophotometry, $V$ band fluxes, $v \sin i$ values,
and \hip\/ distances, while the comparison values are based on
$uvby\beta$ photometry alone --- although both are ultimately tied to
the \atlas\/ models.  This comparison demonstrates that both approaches
are measuring the same quantities and gives us confidence in our
approach.  It also demonstrates that the calibration of the synthetic
$uvby\beta$ photometry that we derive will be nearly identical to the
one used by NSW.  In addition, it shows minimal systematic differences
between the two methods of calculating physical properties.
  
We can also examine how well our derived abundances agree with previous
determinations.  As FM point out, the abundances we measure result from
features in the UV continua, and these are primarily due to iron group
elements.  One source of iron-group abundances for bright, normal B
stars is the Smith \& Dworetsky (1993) fine-analysis of high-resolution
\iue\/ spectra.  Although we have only 5 stars in common (HD~38899,
HD~147394, HD~172167, HD~193432 and HD~215573), it is encouraging that,
of these, the ones with the largest and smallest Fe abundances found by
Smith \& Dworetsky (HD~147394 and HD~172167, respectively) also have
the lowest and largest [m/H] and that the range is similar, 0.70~dex
for Fe and 0.75~dex for [m/H].  Recently, Hempel \& Holweger (2003)
determined Fe abundances for two of our program stars, HD~21790 and
HD~218045, from a NLTE analysis of optical iron lines.  They found
[Fe/H] values of -0.43 and -0.10, respectively, for a difference of
-0.33.  For the same stars, we find [m/H] of -0.64 and -0.02, for a
difference of -0.62.  Again, the ordering is correct, and the magnitude
of the difference agrees reasonable well, considering the very different
methods used.

In addition to abundances, Smith \& Dworetsky also provide \vturb\/
measurements, but there is no relation between their values and ours.
This is not surprising, given the different methods used to arrive at
this parameter and the fact that range observed in our current sample
barely exceeds the internal errors (see Table 5).

Finally, angular diameter measurements by Hanbury Brown et al.~(1974)
are available for three of our program stars, $\gamma$~Gem,
$\alpha$~Leo and $\alpha$~Lyr.  They obtained limb darkened angular
diameters of $1.39 \pm 0.09$, $1.37 \pm 0.06$ and $3.24 \pm 0.07$~mas,
respectively, for these stars.  Using our derived radii from Table~5
and the \hip\/ distances from Table~1, we determine angular diameters
of $1.43 \pm 0.02$, $1.39 \pm 0.02$ and $3.28 \pm 0.04$~mas for the
same stars, where our major source of error is the uncertainties in the
\hip\/ distances.  We see that all of the available measurements are
consistent to less than one $\sigma$.

The comparisons presented in this section show that our calibration (which 
relies only on UV spectrophotometry, calibrated $V$ photometry and \hip\/ 
distances), yields physical properties that agree with those found by 
commonly accepted methods.

\subsection{Errors in the derived parameters}\label{syserrors}

While the formal fitting errors in the derived stellar properties
listed in Table~\ref{tabPARMS} are quite small, we recognize that the
actual errors in the physical properties may be considerably larger,
due to systematic effects.  To properly evaluate such effects, we need
a determination of the stellar properties based on an independent data
set.  The \teff\/ and \logg\/ values determined from $uvby\beta$
photometry form such a set.  As noted earlier, the $uvby\beta$ and
\hip\/-constrained data sets are completely independent of one another,
making their comparison free of bias.  
To perform the comparison, we
simply examine the standard deviation of the differences between the
parameters derived from the two data sets and then subtract,
quadratically, the expected error for the $uvby\beta$ determinations.
The result is an independent estimate of the errors in the
\hip\/-constrained properties.

To proceed, we must first determine the expected errors in the \teff\/
and \logg\/ values derived from the $uvby\beta$ data.  This is done by
performing a Monte Carlo simulation of the sort described in
\S~\ref{secERR}.  Using the values listed in Table~\ref{tabSCAT} for
the errors in the $uvby\beta$ indices, we ran 100 trials for each star
with the random errors added to the photometric indices and then
calculated new values of \teff\/ and \logg\/ with the Napiwotzki
program to obtain the variances for each star.  We then calculated the
RMS error expected from random uncertainties in the $uvby\beta$
photometry.  The results are $\sigma(\log T_{{\rm eff}})_{uvby\beta} =
0.0072$ and $\sigma(\log g)_{uvby\beta} = 0.127$.

We now have the data required to evaluate our errors, and we begin with
\teff.  The standard deviation of the differences between the \hip\/
and $uvby\beta$ determinations is $\sigma(\Delta \log$\teff$) = 0.0090$
and, thus, the predicted error for the \hip-based temperatures is
$\sqrt{0.0090^2 -0.0072^2} = 0.0054$.  The RMS mean of the formal
errors listed in Table~\ref{tabPARMS} is 0.0048 and so we see that the
derived error is only slightly larger than the internal error (1.3\% in
\teff\/ as opposed to 1.1\%).  Thus, we suspect that the \teff\/ are
relatively free of systematics and that the formal fitting errors are
close to the actual errors, and somewhat smaller than the
errors in the \teff\/ determined from $uvby\beta$ photometry (1.3\% vs.
1.7\%).

Turning now to the surface gravity, we find that the standard deviation
of the differences of all the data in Figure 10 is $\sigma(\Delta \log
g) = 0.132$, which implies an expected error in our \hip\/-based
results of $\sqrt{0.132^2 -0.127^2} = 0.036$.  This is actually smaller 
than the RMS of the formal errors for $\log g(Spec)$ in
Table~\ref{tabPARMS}, which is 0.063.  As with \teff, the comparison
suggests that our determinations of $\log g(Spec)$ are free of strong
systematics and that the formal fitting errors are consistent with the
actual errors, and considerably smaller than the errors in the \logg\/
determined from $uvby\beta$ photometry.

The consistency between our $\log g(Spec)$ values and those based on
$uvby\beta$ data provides an important validation for our correction of
the Newtonian gravities for the effects of rotation (see Eq.
\ref{gspec} in \S 2).  The impact that this correction has had on our
results is illustrated in Figure \ref{scatter} where we plot the
residuals between the $uvby\beta$-based gravities and our values for
$\log g(Newton)$ and $\log g(Spec)$, both as a function of $v \sin i$.
The top panel shows that, had we {\it not} corrected the Newtonian
gravities for the effects of rotation, the overall scatter between our
results and the $uvby\beta$-based results would have increased (from
$\pm$0.132 to $\pm$0.159) and a strong systematic effect would have
been introduced into our results, whose roots can be traced all the way
down to $v \sin i \simeq 200$ km/sec.  In our sample, the star with the
largest positive residual (i.e., $\log g - \log g(uvby\beta)$) is HD
225132. (This star, located at $v \sin i = 190$ \kms, lies outside the
bounds of the top panel of Figure \ref{scatter} but stands out clearly
in the lower panel.)  If this discrepancy were a simple rotational
effect, it would require a $v \sin i$ value of greater than 350 km/sec
for our first-order correction to reconcile $\log g(Spec)$ with $\log
g(uvby\beta)$. However, the observed value of 190 km/sec seems
well-determined and is clearly too slow to explain the discrepancy in a
simple way.  It could be that this star is an extremely rapidly
rotating object which is viewed close to pole on, and is sufficiently
distorted to make our correction invalid.  Such a situation might be
revealed through detailed study of its absorption line profiles. This
star deserves further attention.

Our approach should not be confused with the direct application of models 
for rotating stars as applied by Wenske \& Sch\"{o}nberner (1993).  They 
derived \teff\/ and \logg\/ for stars from $uvby\beta$ photometry and then used Collins et al.\ (1991) models and observed \vsini\/ 
measurements to ``correct'' both \teff\/ and \logg\/ for the most probable 
effects of rotation and to infer the \teff\/ and \logg\/ of an equivalent 
{\em non-rotating} star of the same mass.  Our approach is more direct.  
To begin, recall that we discarded Be stars from our sample, so we have 
probably eliminated stars rotating within 75\% of their critical velocity 
(see Massa 1975).  We then obtain a \teff\/ and $E(B-V)$ by fitting the 
observed UV SED and $V$ photometry to a model, assuming $\log g = 4.0$ for 
the initial iteration.  With this estimate for \teff\/ and the \hip\/ 
distance, we calculate the stellar radius.  We then determine the mass 
of the star (and hence its Newtonian gravity) from its position on a 
theoretical HRD expressed in terms of \teff\/ and $R/R_\odot$.  This 
\logg\/ is used to redetermine the \teff\/ and $E(B-V)$.  We then derive 
a new mass and \logg, iterating to consistency.  Throughout the iteration 
process, \teff\/ changes by typically less than 50~K.  The result of our 
fits are an actual \teff\/ (not the temperature of an equivalent, 
non-rotating star of the same mass) which must agree with the integrated 
surface flux of the star.  Similarly, the \logg\/ we derive is a 
Newtonian surface gravity, determined directly from the stellar mass and 
radius.  However, the \logg\/ which corresponds to the $\beta$ photometry 
is a measure of the mean atmospheric pressure, and not the Newtonian 
gravity.   Consequently, we must adjust the Newtonian surface gravity to 
that expected for a rotating star -- otherwise the predicted $\beta$ 
index will not agree with the observed value.  Our procedure is verified 
by the positions of the most rapidly rotating stars in the photometric 
transformation diagrams (Figs.~4--7).  Their points are circled in these 
diagrams and are generally unremarkable, particularly in the $\beta$ 
diagram, suggesting that the Repolust et al.\ (2004) corrections are 
reasonable.  Note, however, that a few of the rapidly rotating stars are 
outliers in the $c1$ and the $(U-B)_{Geneva}$ diagrams, even though their 
corrected $\beta$ indices (which are most sensitive to surface gravity) 
are normal.  This may indicate that the SEDs of the most rapidly rotating 
stars are affected by rotation in other ways.  One possible cause is the 
expected non-uniform temperature distribution over the surface of a 
rapidly rotating star, an effect not included in the Repolust et al.\ 
corrections. It would take a more focused and detailed investigation to isolate such effects

Notice also that there are several stars with large negative residuals
in Figure~\ref{scatter}, particularly HD~153808, HD~188665, and
HD~199081, for which the residuals are $<$ -0.2 dex in both panels of the
figure.  Two of these, (HD~153808 and HD~199081, shown with open
symbols in Fig.~\ref{scatter}) are double-lined spectroscopic binaries
(``SB2s'').  Since our analysis has treated these systems as single
objects, the combination of their observed brightness and the distance
constraint imposed by the \hip\/ data yielded a radius larger than
either member of the binary, resulting in an underestimate of the
surface gravity (see Figure \ref{figHRD}).  In the case of two
identical stars, the result is a simple factor-of-2 error, with the
\hip-based gravities being $\sim$0.3 dex too small, comparable to the
observed discrepancy for HD~153808 and HD~199081.  The third star, HD
188665, has a similar discrepancy, although it is not known to be a
binary.  The data in Figure \ref{scatter} might be evidence for the
binarity of this object.  Note that this binary-induced error in $\log
g$ is peculiar to our \hip-based  approach.  Spectrospcopic
determinations are immune to the effect and will yield a value of $\log
g$ which is a weighted mean of the two binary members, depending on
their relative contributions to the light.

The total errors affecting [m/H] and \vturb\/ are more difficult to
assess.  It is clear from the previous section that the our fitting
procedure identifies specific stars which are known to have extreme
abundance anomalies.  However, we lack the large, uniformly-analyzed
sample of independently determined abundances which is needed to obtain
a quantitative estimate of the total error affecting our metalicity
measurements.  As for \vturb, the question of errors is nearly
irrelevant for the current sample.  By confining our sample to nearby,
low luminosity stars, the maximum value of \vturb\/ for any star in our
sample is only 2.1~\kms, and typical errors are $0.5 - 1.0$~\kms.  As a
result, it is problematic to identify any systematic dependencies
involving \vturb\/ from the current sample.  In the future, we may be
able to assess the accuracy of [m/H] and \vturb\/ indirectly by, for
example, examining the scatter of [m/H] within a cluster, or the
dependency of \vturb\/ upon luminosity is a larger sample of stars.
However, until such samples are available, we must consider the values
listed in Table~\ref{tabPARMS} as lower limits to the actual errors for
these parameters.

In summary, we see that the our estimates for \teff\/ are free of
systematics and have random errors that are somewhat smaller than the
errors affecting temperatures derived from optical photometry alone.
We have also seen that our \hip-based approach has yielded accurate
values of $\log g(Spec)$ with small and well-determined random errors.
On the other hand, our analysis has highlighted an important systematic
effect which may affect many analyses.  Namely, that
spectroscopically-based surface gravity determinations (e.g., from
$\beta$ photometry or from line profile analysis) may systematically
underestimate the actual Newtonian gravity (i.e., $GM/R^2$) of a star,
in the presence of even moderate rotation (i.e., $v \sin i > 200$
km/sec).  The most obvious area of concern is when
spectroscopically-based gravities are used to place stars in
theoretical HR Diagrams for comparison with stellar structure and
evolution models, as discussed by Herrero et al. (1992).  Knowledge of a
star's rotation would appear to be an essential piece of information
for such studies.

\subsection{Applying the calibration}

In our own future analyses, the most common application of the
calibrated synthetic photometry will be to model the SEDs of stars
(both reddened and unreddened) by combining \iue\/ UV spectrophotometry
and an ensemble of available ground-based photometry.  This combination
provides powerful constraints on SED models, including both the
intrinsic properties of the stars and the effects of interstellar
extinction. In particular, the UV data provide exclusive information on
\vturb, [m/H], and the shape of the UV extinction law.  The combined UV
and optical/near-IR photometry present a very long wavelength baseline
for the determination of \teff.  In addition, the optical and near-IR
continuum data provide access to the Balmer Jump (a useful temperature
and gravity diagnostic) and the important extinction parameters
$E(B-V)$ and $A(V)$.  Finally, $\beta$ photometry is extremely
sensitive to luminosity (Crawford 1958) or, equivalently, \logg.  When
a full suite of photometry is available, there is considerable
redundancy.  However, this redundancy reduces the effects of random
errors in the individual measurements, and allows the identification of
the occasional pathological data point.

When fitting a UV/optical/near-IR SED, we must decide how to weight the
various data sets in the least-squares procedure. Clearly, if we simply
use the observational errors, this would assign most of the weight to
the UV, since it contains so many points.  However, this is
undesirable since the stellar and interstellar information is spread
throughout the spectrum. Given this distribution of information, we
decided to divide the fitting weights into three groups: one for the
UV, one for the optical/near-IR continuum photometry, and one for the $\beta$
photometry.  After considerable experimentation, we adopted the simple
procedure of assigning identical total weight to each group, and
distributing the weights within a group according to the observational
errors.  This balance between the UV and the optical/near-IR is
intuitively appealing, although the great weight given to a single
photometric index, $\beta$, merits comment. We have adopted this
procedure essentially to insure that the \logg\/ results of the fitting
procedure --- in the absence of a \hip\/ constraint which would
otherwise select the appropriate gravity --- be consistent with
photometric results based on $uvby\beta$ photometry, as discussed in
the previous section.

To test the above approach, we applied the SED fitting technique to our
45-star sample and calculated errors via the Monte Carlo approach
described in \S~4.1, except that random realizations of the input
photometric indices --- based on their expected errors (see Table 7)
--- were also included.  Next, we examined the differences between the
\hip\/-based and the new SED-based properties for \teff, [m/H], and
\vturb, and between the $uvby\beta$-based and SED-based values for
\logg. The RMS errors for these samples are $\sigma(\log$\teff$)_{hip}
= 0.0048$, $\sigma($[{\rm m/H}]$)_{hip} = 0.138$,
$\sigma($\vturb$)_{hip} = 0.644$, and $\sigma(\log$\teff$)_{SED} =
0.0033$, $\sigma($[{\rm m/H}]$)_{SED} = 0.111$, $\sigma($\vturb$)_{SED}
= 0.797$, for the \hip\/ and SED based parameters respectively.  The
means and standard deviations of the differences between these
properties are: $\Delta \log$\teff$ = -0.0013 \pm 0.0066$;
$\Delta$[m/H]$= 0.019 \pm 0.132$; and $\Delta$\vturb$= -0.14 \pm
0.579$.  For totally uncorrelated data, we would expect $\sigma(\Delta
x) \simeq  \sqrt{\sigma(x)_{hip}^2 + \sigma(x)_{SED}^2}$, while for
completely correlated data, we expect $\sigma(\Delta x) \simeq 0$.
 For [m/H] and \vturb, we find $\sigma(\Delta x) \simeq \sigma(x)_{hip}
\simeq \sigma(x)_{SED}$, implying that errors in these properties are
strongly correlated.  This is because these measurements depend almost
exclusively on the UV SED (which is common to both data sets), and only
weakly on the optical/near-IR data.  For \teff, we find that
$\sigma(\Delta$\teff)\/ and $\sqrt{ 0.0048^2 + 0.0033^2} = 0.0058$, are
comparable, suggesting that the \teff\/ values derived from the SEDs
are strongly influenced by the optical/near-IR data.  This is
reasonable, since \teff\/ determinations are sensitive to the total
wavelength baseline, and the SED values incorporate significant,
uncorrelated data.  Finally, we compared our SED-based surface
gravities with the $uvby\beta$-based values discussed in \S 7.2, since
our goal is to make the two as consistent as possible.  The results
are:  $\sigma($\logg$)_{uvby\beta} = 0.117$, $\sigma($\logg$)_{SED} =
0.106$, and $\Delta(\log g) = \log g_{SED} - \log g_{uvby\beta} =
-0.017 \pm 0.061$ for the entire sample.  This time, we find that the
standard deviation in $\Delta(\log g)$ is much less than {\em either}
$\sigma(\log g)_{SED}$ or $\sigma(\log g)_{uvby\beta}$, indicating a
very high degree of correlation, as desired.

Thus, although the details of the fitting procedure are somewhat
subjective, the results of this section demonstrate that we are  able
to achieve excellent agreement between the physical properties
determined from the SED fitting (without the \hip\/ constraint) and the
other techniques.

\section{Summary and Future Directions}

To summarize, we have derived a calibration of synthetic optical and
NIR photometry from \atlas\/ model atmospheres which is based on \iue\/
UV spectrophotometry, $V$ band fluxes and \hip\/ distances.  The result
is a set of calibrated synthetic photometry which is consistent from
the UV to NIR.  We have also shown that spectroscopically-based
determinations of \logg\/ --- such as derived from $\beta$ photometry
--- for stars with $v \sin i \geq 200$~km~s$^{-1}$ are systematically
lower than the true Newtonian gravities, $GM/R^2$.

The ability to predict the intrinsic UV continuum of a star and to
derive [m/H] and \vturb\/ will allow us to pursue several important
applications.  Some of these are:

\begin{enumerate}
  \item {\bf Large study of unreddened stars:}  We have identified
   approximately 150 lightly reddened early A and B stars with high
   quality \iue\/ data and optical photometry.  This sample will be
   used to search for systematic effects among the derived physical
   properties.  It will also be used to extend the present results to
   hotter stars, in order to better define the range of applicability of
   the our approach.  In particular, the availability of a large
   sample of stars, with uniformly determined \vturb\/ values,  will
   enable us to examine the dependence this poorly understood parameter
   on other factors, such as \teff, \logg, mass loss, \vsini, etc.

  \item {\bf UV reddening:}  Because the models provide a precise
   intrinsic stellar continuum, we can use our approach to determine the
   UV extinction affecting a star without appealing to the subjective
   process of selecting an unreddened spectral match.  FM provided an
   example of an extinction curve derived by fitting a model to a
   moderately reddened star.  More recently, Massa \& Fitzpatrick (2004)
   demonstrated how this same approach can be used to derive extinction
   curves from lightly reddened stars, enabling studies of local and
   Galactic halo dust.

  \item {\bf Open Clusters:}  Fitzpatrick \& Massa (1990) list 5 open
    clusters which contain
    several main sequence B stars and additional examples exist in the
    \iue\/ archives.  By fitting the B star members of open clusters,
    we will be able to: 1)  determine the small scale uniformity of UV
    extinction much more precisely than previously possible, 2) derive
    $R(V)$ values, needed to obtain accurate unreddened magnitudes, 3)
    calculate precise temperatures and gravities, needed to construct
    theoretical HRDs, 5) derive metalicities, to study the uniformity
    of cluster composition, and 6) determine the unreddened continua of
    cluster stars with unknown continua (Massa et al.\ 1985).

  \item {\bf The distribution of metallicity:} The ability to determine
    accurate metallicities for virtually every main sequence B and early-A
    star observed by
    \iue\/ will enable us to search for systematic patterns in the
    distribution of metalicity in cluster and field A and B stars.

  \item {\bf Far UV fluxes:} Using the intrinsic fluxes determined from
    \iue\/ and optical photometry, we will be able to test the validity of
    the models in the FUV by comparing them to {\it FUSE}\/  observations
    of lightly reddened and reddened B stars.  This will be a challenging
    test of the models, since the FUV is rich in spectral lines.  If the
    comparisons for lightly reddened stars are favorable, we will have
    the confidence needed to derive accurate FUV extinction curves for
    reddened B stars.
\end{enumerate}

Clearly, the approach outlined in this paper opens the door for several
types of fundamental research on both stars and interstellar dust.

\begin{acknowledgments}
We are very grateful for a detailed and thoughtful review of this paper
by the referee, Dr. R. Napiwotzki.  E.F. acknowledges support from NASA
grant NAG5-12137 and thanks Bernard Nicolet for making the Geneva
photometry passbands available.  D.M.  acknowledges support from NASA
grant NNG04EC01P. Some of the data presented in this paper were
obtained from the Multimission Archive at the Space Telescope Science
Institute (MAST).  STScI is operated by the Association of Universities
for Research in Astronomy, Inc., under NASA contract NAS5-26555.
Support for MAST for non-HST data is provided by the NASA Office of
Space Science via grant NAG5-7584 and by other grants and contracts.
This publication also makes use of data products from the Two Micron
All Sky Survey, which is a joint project of the University of
Massachusetts and the Infrared Processing and Analysis
Center/California Institute of Technology, funded by the National
Aeronautics and Space Administration and the National Science
Foundation.
\end{acknowledgments}

%------------------------------------------------------------------------%
% References                                                             %
%------------------------------------------------------------------------%

\bibliographystyle{apj}

%------------------------------------------------------------------------%
% TABLES                                                                 %
%------------------------------------------------------------------------%
%%%%%%%%%%%%%%%%%%%%%%%%%%%%%%% TABLE 1 %%%%%%%%%%%%%%%%%%%%%%%%%%%%%%%%%%

\begin{deluxetable}{rlcclc} 
\tabletypesize{\footnotesize}
\tablewidth{0pc} 
\tablecaption{Basic Data for Program Stars
\label{tabSTARS}} 
\tablehead{ 
\colhead{HD}                          &  
\colhead{Star}                        &  
\colhead{\it Hipparcos}               &  
\colhead{$v \sin i$\tablenotemark{b}} &  
\colhead{Spectral}                    &  
\colhead{Spectral Type}                        \\

\colhead{Number}                      &  
\colhead{Name}                        &  
\colhead{Distance\tablenotemark{a}}   &  
\colhead{(km s$^{-1}$)}               &  
\colhead{Type}                        &  
\colhead{Reference\tablenotemark{c}}       }
\startdata     
     886 &  $\gamma$ Peg   &  $102.1\pm8.2\phn$       &  \phn\phn5&    B2 IV     & 6     \\
    9132 &   48 Cet        &  $\phn67.9\pm3.4\phn$    &  \phn20   &    A1 Va     & 3    \\
   10250 &  42 Cas         &  $\phn86.1\pm4.3\phn$    &  125      &    B9 V      & 2    \\
   21790 &  17 Eri         &  $116.7\pm10.5$          &  \phn85   &    B9 III    & 4    \\
   29646 &  \nodata        &  $102.7\pm9.2\phn$       &  120      &    A1 IV     & 3    \\
   32630 &  $\eta$ Aur     &  $\phn67.2\pm3.4\phn$    &  125      &    B3 V      & 6     \\
   38899 &  134 Tau        &  $\phn83.3\pm5.8\phn$    &  \phn25   &    B9 IV     & 4    \\
   45557 &  \nodata        &  $\phn88.0\pm3.5\phn$    &  \nodata  &    A0 V      & 1     \\
   47105 &  $\gamma$ Gem   &  $\phn 32.1\pm2.2\phn$   &  \phn30   &    A1 IVs    & 3    \\
   58142 &  21 Lyn         &  $\phn76.3\pm4.6\phn$    &  \phn10   &    A0mA1 IV  & 3    \\
   61831 &  \nodata        &  $176.1\pm14.1$          &  180      &    B2.5 V    & 5     \\
   66591 &  \nodata        &  $166.1\pm13.3$          &  \phn50   &    B3 V      & 5     \\
   77002 &  \nodata        &  $190.5\pm17.1$          &  \phn50   &    B2 IV-V   & 5     \\
   79447 &  \nodata        &  $153.1\pm10.7$          &  \phn10   &    B3 III    & 5     \\
   87901 &  $\alpha$ Leo   &  $\phn23.8\pm0.5\phn$    &  330      &    B8 Vnn    & 4    \\
   93194 &  \nodata        &  $148.4\pm10.4$          &  295      &    B4 IVn    & 5     \\
   98664 &  $\sigma$ Leo   &  $\phn65.6\pm3.3\phn$    &  \phn65   &    A0 III+   & 3    \\
  103287 &  $\gamma$ UMa   &  $\phn 25.6\pm0.5\phn$   &  170      &    A0 Van    & 3    \\
  106911 &  $\beta$ Cha    &  $\phn 83.0\pm3.3\phn$   &  260      &    B5 Vn     & 5    \\
  115823 &  \nodata        &  $121.4\pm9.7\phn$       &  \phn80   &    B6 V      & 5     \\
  121263 &  $\zeta$ Cen    &  $117.9\pm10.6$          &  225      &    B2.5 IV   & 5     \\
  121743 &  $\phi$ Cen     &  $142.7\pm14.3$          &  115      &    B2 IV     & 5     \\
  121790 &  $\nu^1$ Cen    &  $128.0\pm11.5$          &  150      &    B2 IV-V   & 5     \\
  125238 &  $\iota$ Lup    &  $107.9\pm8.6\phn$       &  235      &    B2.5 IV   & 5     \\
  129116 &  \nodata        &  $\phn93.5\pm6.5\phn$    &  185      &    B3 V      & 5     \\
  147394 &  $\tau$ Her     &  $\phn 96.4\pm4.8\phn$   &  \phn30   &    B5 IV     & 6     \\
  153808 &  $\epsilon$ Her &  $\phn49.9\pm1.5\phn$    &  \phn85   &    A0 IV+    & 3    \\
  158094 &  $\delta$ Ara   &  $\phn57.4\pm2.3\phn$    &  255      &    B8 IVn    & 4    \\
  160762 &  $\iota$ Her    &  $152.0\pm 13.7$         &  \phn10   &    B3 IV     & 6     \\
  172167 &  $\alpha$ Lyr   &  $\phn\phn7.8\pm0.0\phn$ &  \phn15   &    A0 Va     & 3    \\
  177724 &  $\zeta$ Aql    &  $\phn25.5\pm0.5\phn$    &  345      &    A0 Vann   & 3    \\
  182255 &  3 Vul          &  $123.5\pm9.9\phn$       &  \phn40   &    B6 III    & 6     \\
  188665 &  23 Cyg         &  $195.7\pm19.6$          &  145      &    B5 V      & 6     \\
  192907 &  $\kappa$ Cep   &  $100.3\pm5.0\phn$       &  \phn20   &    B9 IV+    & 4    \\
  193432 &  $\nu$ Cap      &  $\phn83.5\pm6.7\phn$    &  \phn15   &    B9.5 Va+  & 3    \\
  196867 &  $\alpha$ Del   &  $\phn73.8\pm3.7\phn$    &  155      &    B9 IV     & 4    \\
  199081 &  57 Cyg         &  $153.6\pm13.8$          &  \phn85   &    B5 V      & 6     \\
  201908 &  \nodata        &  $126.4\pm7.6\phn$       &  225      &    B8 Vn     & 2    \\
  210419 &  \nodata        &  $111.1\pm10.0$          &  370      &    A0 IVnn   & 3    \\
  214923 &  $\zeta$ Peg    &  $\phn 63.9\pm3.2\phn$   &  185      &    B8.5 III  & 4    \\
  215573 &  $\xi$ Oct      &  $136.1\pm8.2\phn$       &  \phn50   &    B6 IV     & 5     \\
  218045 &  $\alpha$ Peg   &  $\phn42.8\pm1.3\phn$    &  150      &    A0 III-IV & 3    \\
  222439 &  $\kappa$ And   &  $\phn52.0\pm1.6\phn$    &  190      &    B9 IVn    & 4    \\
  222661 &  $\omega^2$ Aqr &  $\phn47.3\pm1.9\phn$    &  145      &    B9.5 IV   & 3    \\
  225132 &  2 Cet          &  $\phn69.9\pm4.2\phn$    &  190      &    B9 IVn    & 4    \\
\enddata
\tablenotetext{a}{From the {\it Hipparcos} Catalog: Perryman et~al. 1997}
\tablenotetext{b}{$v \sin i$ values from Glebocki \& Stawikowski
2000}
\tablenotetext{c}{Spectral type references: (1) Hoffleit 1982; (2) 
Cowley et~al. 1969; (3) Gray \& Garrison 1987; (4) Garrison \& Gray  
1987; (5) Hilter, Garrison, \& Schild 1969; (6) Lesh 1968} 
\end{deluxetable}
\clearpage

%%%%%%%%%%%%%%%%%%%%%%%%%%%%%%% TABLE 2 %%%%%%%%%%%%%%%%%%%%%%%%%%%%%%%%%%

\begin{deluxetable}{lrrrrrrrrrrrr} 
\tabletypesize{\scriptsize}
\tablewidth{0pc} 
\tablecaption{Johnson and Str\"{o}mgren Photometry \label{tabPHOT}}
\tablehead{ 
\colhead{Star}     &  
\colhead{$V$}      & 
\colhead{$B-V$}    & 
\colhead{$U-B$}    & 
\colhead{$V-R$}    & 
\colhead{$R-I$}    & 
\colhead{$J$}      & 
\colhead{$H$}      & 
\colhead{$K$}      & 
\colhead{$b-y$}    & 
\colhead{$m1$}     & 
\colhead{$c1$}     & 
\colhead{$\beta$}   }
\startdata
    HD 886 & $ 2.83$ & $-0.22$ & $-0.86$ & $-0.10$ & $-0.18$ & $ 3.34$ & \nodata & $ 3.52$ & $-0.11$ & $ 0.09$ & $ 0.12$ & $ 2.627$ \\
   HD 9132 & $ 5.12$ & $ 0.02$ & $ 0.05$ & \nodata & \nodata & \nodata & \nodata & \nodata & $ 0.01$ & $ 0.16$ & $ 1.09$ & $ 2.901$ \\
  HD 10250 & $ 5.18$ & $-0.04$ & \nodata & \nodata & \nodata & \nodata & \nodata & \nodata & $-0.02$ & $ 0.15$ & $ 1.02$ & \nodata \\
  HD 21790 & $ 4.73$ & $-0.09$ & $-0.26$ & $-0.01$ & $-0.08$ & \nodata & \nodata & \nodata & $-0.04$ & $ 0.11$ & $ 0.83$ & $ 2.761$ \\
  HD 29646 & $ 5.74$ & $-0.02$ & $ 0.03$ & \nodata & \nodata & \nodata & \nodata & \nodata & $-0.01$ & $ 0.17$ & $ 1.04$ & $ 2.909$ \\
  HD 32630 & $ 3.17$ & $-0.18$ & $-0.67$ & $-0.05$ & $-0.18$ & $ 3.55$ & $ 3.61$ & $ 3.68$ & $-0.09$ & $ 0.10$ & $ 0.32$ & $ 2.683$ \\
  HD 38899 & $ 4.90$ & $-0.07$ & $-0.17$ & $ 0.02$ & $-0.08$ & \nodata & \nodata & \nodata & $-0.03$ & $ 0.14$ & $ 0.91$ & $ 2.825$ \\
  HD 45557 & $ 5.79$ & $ 0.00$ & \nodata & \nodata & \nodata & \nodata & \nodata & \nodata & $-0.01$ & $ 0.17$ & $ 1.03$ & $ 2.891$ \\
  HD 47105 & $ 1.93$ & $ 0.00$ & $ 0.05$ & $ 0.05$ & $-0.01$ & $ 1.90$ & $ 1.88$ & $ 1.87$ & $ 0.01$ & $ 0.15$ & $ 1.19$ & $ 2.865$ \\
  HD 58142 & $ 4.61$ & $-0.01$ & $-0.01$ & $-0.01$ & $-0.05$ & \nodata & \nodata & \nodata & $ 0.00$ & $ 0.14$ & $ 1.12$ & $ 2.875$ \\
  HD 61831 & $ 4.84$ & $-0.19$ & $-0.65$ & $-0.08$ & $-0.16$ & \nodata & \nodata & \nodata & $-0.08$ & $ 0.11$ & $ 0.30$ & $ 2.669$ \\
  HD 66591 & $ 4.81$ & $-0.17$ & $-0.62$ & $-0.04$ & $-0.20$ & \nodata & \nodata & \nodata & $-0.08$ & $ 0.10$ & $ 0.31$ & $ 2.678$ \\
  HD 77002 & $ 4.91$ & $-0.18$ & $-0.74$ & $-0.09$ & $-0.24$ & \nodata & \nodata & \nodata & $-0.09$ & $ 0.10$ & $ 0.21$ & $ 2.659$ \\
  HD 79447 & $ 3.97$ & $-0.19$ & $-0.67$ & $-0.05$ & $-0.18$ & \nodata & \nodata & \nodata & $-0.08$ & $ 0.10$ & $ 0.30$ & $ 2.659$ \\
  HD 87901 & $ 1.36$ & $-0.11$ & $-0.36$ & $-0.02$ & $-0.10$ & $ 1.56$ & $ 1.58$ & $ 1.62$ & $-0.04$ & $ 0.10$ & $ 0.71$ & $ 2.723$ \\
  HD 93194 & $ 4.82$ & $-0.14$ & $-0.62$ & $-0.01$ & $-0.14$ & \nodata & \nodata & \nodata & $-0.06$ & $ 0.10$ & $ 0.36$ & $ 2.668$ \\
  HD 98664 & $ 4.04$ & $-0.06$ & $-0.11$ & \nodata & \nodata & \nodata & \nodata & \nodata & $-0.02$ & $ 0.13$ & $ 1.01$ & $ 2.827$ \\
 HD 103287 & $ 2.44$ & $ 0.00$ & $ 0.01$ & $ 0.00$ & $-0.04$ & $ 2.40$ & \nodata & $ 2.37$ & $ 0.01$ & $ 0.16$ & $ 1.11$ & $ 2.885$ \\
 HD 106911 & $ 4.24$ & $-0.13$ & $-0.52$ & $-0.02$ & $-0.11$ & \nodata & \nodata & \nodata & $-0.05$ & $ 0.10$ & $ 0.45$ & $ 2.710$ \\
 HD 115823 & $ 5.47$ & $-0.13$ & $-0.53$ & $-0.08$ & $-0.15$ & \nodata & \nodata & \nodata & $-0.07$ & $ 0.12$ & $ 0.46$ & $ 2.734$ \\
 HD 121263 & $ 2.54$ & $-0.23$ & $-0.90$ & $-0.14$ & $-0.20$ & $ 3.09$ & \nodata & $ 3.29$ & $-0.11$ & $ 0.08$ & $ 0.05$ & $ 2.619$ \\
 HD 121743 & $ 3.82$ & $-0.22$ & $-0.83$ & $-0.13$ & $-0.22$ & \nodata & \nodata & \nodata & $-0.10$ & $ 0.08$ & $ 0.14$ & $ 2.635$ \\
 HD 121790 & $ 3.86$ & $-0.21$ & $-0.80$ & $-0.14$ & $-0.21$ & $ 4.37$ & $ 4.48$ & $ 4.52$ & $-0.10$ & $ 0.09$ & $ 0.16$ & $ 2.640$ \\
 HD 125238 & $ 3.55$ & $-0.18$ & $-0.72$ & $-0.10$ & $-0.14$ & \nodata & \nodata & \nodata & $-0.08$ & $ 0.09$ & $ 0.25$ & $ 2.650$ \\
 HD 129116 & $ 3.99$ & $-0.17$ & $-0.69$ & $-0.11$ & $-0.15$ & \nodata & \nodata & \nodata & $-0.08$ & $ 0.09$ & $ 0.25$ & $ 2.672$ \\
 HD 147394 & $ 3.90$ & $-0.15$ & $-0.56$ & $-0.09$ & $-0.17$ & \nodata & \nodata & \nodata & $-0.06$ & $ 0.09$ & $ 0.44$ & $ 2.702$ \\
 HD 153808 & $ 3.92$ & $-0.02$ & $-0.10$ & $-0.01$ & $-0.04$ & \nodata & \nodata & \nodata & $-0.00$ & $ 0.15$ & $ 0.92$ & $ 2.861$ \\
 HD 158094 & $ 3.60$ & $-0.10$ & $-0.30$ & \nodata & \nodata & \nodata & \nodata & \nodata & $-0.04$ & $ 0.10$ & $ 0.78$ & $ 2.772$ \\
 HD 160762 & $ 3.80$ & $-0.18$ & $-0.70$ & $-0.10$ & $-0.17$ & \nodata & \nodata & \nodata & $-0.06$ & $ 0.08$ & $ 0.29$ & $ 2.661$ \\
 HD 172167 & $ 0.03$ & $-0.00$ & $-0.01$ & $-0.04$ & $-0.02$ & $ 0.02$ & $ 0.00$ & $ 0.02$ & $ 0.00$ & $ 0.16$ & $ 1.09$ & $ 2.903$ \\
 HD 177724 & $ 2.99$ & $ 0.01$ & $ 0.01$ & $ 0.01$ & $ 0.00$ & \nodata & \nodata & \nodata & $ 0.01$ & $ 0.15$ & $ 1.08$ & $ 2.875$ \\
 HD 182255 & $ 5.19$ & $-0.12$ & $-0.52$ & $-0.04$ & $-0.16$ & \nodata & \nodata & \nodata & $-0.05$ & $ 0.11$ & $ 0.49$ & $ 2.736$ \\
 HD 188665 & $ 5.14$ & $-0.13$ & $-0.55$ & \nodata & \nodata & \nodata & \nodata & \nodata & $-0.06$ & $ 0.10$ & $ 0.45$ & $ 2.715$ \\
 HD 192907 & $ 4.38$ & $-0.05$ & $-0.11$ & $-0.02$ & $-0.06$ & \nodata & \nodata & \nodata & $-0.02$ & $ 0.13$ & $ 1.03$ & $ 2.825$ \\
 HD 193432 & $ 4.75$ & $-0.05$ & $-0.10$ & $ 0.01$ & $-0.06$ & $ 4.84$ & $ 4.90$ & $ 4.85$ & $-0.02$ & $ 0.13$ & $ 1.01$ & $ 2.852$ \\
 HD 196867 & $ 3.77$ & $-0.06$ & $-0.21$ & $ 0.00$ & $-0.04$ & \nodata & \nodata & \nodata & $-0.02$ & $ 0.12$ & $ 0.89$ & $ 2.796$ \\
 HD 199081 & $ 4.77$ & $-0.14$ & $-0.58$ & $-0.07$ & $-0.13$ & $ 5.09$ & $ 5.12$ & $ 5.16$ & $-0.05$ & $ 0.10$ & $ 0.40$ & $ 2.713$ \\
 HD 201908 & $ 5.91$ & $-0.07$ & $-0.24$ & \nodata & \nodata & \nodata & \nodata & \nodata & $-0.04$ & $ 0.13$ & $ 0.80$ & $ 2.805$ \\
 HD 210419 & $ 6.26$ & $-0.01$ & $-0.07$ & \nodata & \nodata & \nodata & \nodata & \nodata & $-0.00$ & $ 0.14$ & $ 1.07$ & $ 2.842$ \\
 HD 214923 & $ 3.40$ & $-0.09$ & $-0.24$ & $-0.04$ & $-0.07$ & \nodata & \nodata & \nodata & $-0.04$ & $ 0.11$ & $ 0.87$ & $ 2.768$ \\
 HD 215573 & $ 5.33$ & $-0.15$ & $-0.49$ & \nodata & \nodata & \nodata & \nodata & \nodata & $-0.07$ & $ 0.11$ & $ 0.51$ & $ 2.718$ \\
 HD 218045 & $ 2.48$ & $-0.04$ & $-0.04$ & $ 0.01$ & $-0.03$ & $ 2.52$ & \nodata & $ 2.54$ & $-0.01$ & $ 0.13$ & $ 1.13$ & $ 2.838$ \\
 HD 222439 & $ 4.14$ & $-0.07$ & $-0.23$ & $-0.01$ & $-0.07$ & \nodata & \nodata & \nodata & $-0.04$ & $ 0.13$ & $ 0.83$ & $ 2.833$ \\
 HD 222661 & $ 4.48$ & $-0.04$ & $-0.12$ & $ 0.02$ & $-0.05$ & \nodata & \nodata & \nodata & $-0.02$ & $ 0.14$ & $ 0.92$ & $ 2.870$ \\
 HD 225132 & $ 4.55$ & $-0.05$ & $-0.08$ & $ 0.03$ & $-0.04$ & \nodata & \nodata & \nodata & $-0.01$ & $ 0.12$ & $ 1.02$ & $ 2.791$ \\
\enddata
\tablecomments{The Johnson and Str\"{o}mgren data were collected from
the General Catalog of Photometric Data maintained by Institute of Astronomy 
of the University of Lausanne (Switzerland).}
\end{deluxetable}
\clearpage

%%%%%%%%%%%%%%%%%%%%%%%%%%%%%%% TABLE 3 %%%%%%%%%%%%%%%%%%%%%%%%%%%%%%%%%%

\begin{deluxetable}{lcccccc} 
\tabletypesize{\footnotesize}
\tablewidth{0pc} 
\tablecaption{Geneva Photometry\label{tabGEN}}
\tablehead{ 
\colhead{Star}        &  
\colhead{$U-B$}       & 
\colhead{$V-B$}       & 
\colhead{$B1-B$}      & 
\colhead{$B2-B$}      & 
\colhead{$V1-B$}      & 
\colhead{$G-B$}       }
\startdata
    HD 886 & $ 0.39$ & $ 1.23$ & $ 0.79$ & $ 1.63$ & $ 1.93$ & $ 2.48$\\
   HD 9132 & $ 1.52$ & $ 0.93$ & $ 0.89$ & $ 1.50$ & $ 1.62$ & $ 2.12$\\
  HD 10250 & $ 1.42$ & $ 0.98$ & $ 0.87$ & $ 1.51$ & $ 1.67$ & $ 2.18$\\
  HD 21790 & $ 1.16$ & $ 1.07$ & $ 0.83$ & $ 1.55$ & $ 1.76$ & $ 2.28$\\
  HD 29646 & \nodata & \nodata & \nodata & \nodata & \nodata & \nodata\\
  HD 32630 & $ 0.63$ & $ 1.17$ & $ 0.81$ & $ 1.60$ & $ 1.87$ & $ 2.41$\\
  HD 38899 & $ 1.28$ & $ 1.05$ & $ 0.86$ & $ 1.54$ & $ 1.74$ & $ 2.27$\\
  HD 45557 & $ 1.46$ & $ 0.97$ & $ 0.88$ & $ 1.51$ & $ 1.67$ & $ 2.17$\\
  HD 47105 & $ 1.60$ & $ 0.96$ & $ 0.89$ & $ 1.52$ & $ 1.66$ & $ 2.15$\\
  HD 58142 & $ 1.50$ & $ 0.97$ & $ 0.87$ & $ 1.50$ & $ 1.66$ & $ 2.16$\\
  HD 61831 & $ 0.59$ & $ 1.18$ & $ 0.80$ & $ 1.60$ & $ 1.86$ & $ 2.41$\\
  HD 66591 & $ 0.62$ & $ 1.15$ & $ 0.80$ & $ 1.58$ & $ 1.85$ & $ 2.38$\\
  HD 77002 & $ 0.51$ & $ 1.18$ & $ 0.79$ & $ 1.59$ & $ 1.86$ & $ 2.41$\\
  HD 79447 & $ 0.59$ & $ 1.18$ & $ 0.79$ & $ 1.59$ & $ 1.87$ & $ 2.40$\\
  HD 87901 & $ 1.05$ & $ 1.09$ & $ 0.83$ & $ 1.57$ & $ 1.79$ & $ 2.31$\\
  HD 93194 & $ 0.66$ & $ 1.13$ & $ 0.80$ & $ 1.58$ & $ 1.82$ & $ 2.34$\\
  HD 98664 & $ 1.36$ & $ 1.03$ & $ 0.85$ & $ 1.52$ & $ 1.73$ & $ 2.24$\\
 HD 103287 & $ 1.54$ & $ 0.95$ & $ 0.89$ & $ 1.51$ & $ 1.66$ & $ 2.15$\\
 HD 106911 & $ 0.78$ & $ 1.11$ & $ 0.81$ & $ 1.56$ & $ 1.80$ & $ 2.32$\\
 HD 115823 & $ 0.78$ & $ 1.12$ & $ 0.82$ & $ 1.56$ & $ 1.80$ & $ 2.33$\\
 HD 121263 & $ 0.32$ & $ 1.24$ & $ 0.78$ & $ 1.62$ & $ 1.93$ & $ 2.49$\\
 HD 121743 & $ 0.39$ & $ 1.22$ & $ 0.78$ & $ 1.61$ & $ 1.90$ & $ 2.46$\\
 HD 121790 & $ 0.42$ & $ 1.22$ & $ 0.79$ & $ 1.61$ & $ 1.90$ & $ 2.46$\\
 HD 125238 & $ 0.54$ & $ 1.18$ & $ 0.80$ & $ 1.61$ & $ 1.88$ & $ 2.42$\\
 HD 129116 & $ 0.54$ & $ 1.16$ & $ 0.80$ & $ 1.59$ & $ 1.86$ & $ 2.41$\\
 HD 147394 & $ 0.75$ & $ 1.15$ & $ 0.81$ & $ 1.58$ & $ 1.84$ & $ 2.37$\\
 HD 153808 & $ 1.37$ & $ 0.99$ & $ 0.89$ & $ 1.52$ & $ 1.68$ & $ 2.19$\\
 HD 158094 & $ 1.09$ & $ 1.08$ & $ 0.82$ & $ 1.55$ & $ 1.77$ & $ 2.29$\\
 HD 160762 & $ 0.59$ & $ 1.17$ & $ 0.80$ & $ 1.60$ & $ 1.86$ & $ 2.40$\\
 HD 172167 & $ 1.50$ & $ 0.96$ & $ 0.90$ & $ 1.51$ & $ 1.66$ & $ 2.17$\\
 HD 177724 & $ 1.51$ & $ 0.94$ & $ 0.89$ & $ 1.51$ & $ 1.64$ & $ 2.13$\\
 HD 182255 & $ 0.81$ & $ 1.11$ & $ 0.82$ & $ 1.56$ & $ 1.80$ & $ 2.33$\\
 HD 188665 & $ 0.76$ & $ 1.10$ & $ 0.82$ & $ 1.58$ & $ 1.82$ & $ 2.33$\\
 HD 192907 & $ 1.38$ & $ 1.01$ & $ 0.86$ & $ 1.54$ & $ 1.72$ & $ 2.23$\\
 HD 193432 & $ 1.38$ & $ 1.01$ & $ 0.86$ & $ 1.52$ & $ 1.70$ & $ 2.21$\\
 HD 196867 & $ 1.25$ & $ 1.03$ & $ 0.85$ & $ 1.54$ & $ 1.73$ & $ 2.23$\\
 HD 199081 & $ 0.72$ & $ 1.13$ & $ 0.81$ & $ 1.58$ & $ 1.82$ & $ 2.36$\\
 HD 201908 & \nodata & \nodata & \nodata & \nodata & \nodata & \nodata\\
 HD 210419 & $ 1.46$ & $ 0.97$ & $ 0.87$ & $ 1.51$ & $ 1.66$ & $ 2.15$\\
 HD 214923 & $ 1.20$ & $ 1.07$ & $ 0.83$ & $ 1.56$ & $ 1.76$ & $ 2.28$\\
 HD 215573 & $ 0.82$ & $ 1.13$ & $ 0.81$ & $ 1.57$ & $ 1.81$ & $ 2.35$\\
 HD 218045 & $ 1.50$ & $ 1.01$ & $ 0.87$ & $ 1.54$ & $ 1.72$ & $ 2.22$\\
 HD 222439 & $ 1.18$ & $ 1.04$ & $ 0.85$ & $ 1.53$ & $ 1.74$ & $ 2.26$\\
 HD 222661 & $ 1.30$ & $ 1.01$ & $ 0.88$ & $ 1.52$ & $ 1.71$ & $ 2.22$\\
 HD 225132 & $ 1.37$ & $ 1.01$ & $ 0.85$ & $ 1.53$ & $ 1.71$ & $ 2.21$\\
\enddata
\tablecomments{The Geneva data were collected from
the General Catalog of Photometric Data maintained by Institute of Astronomy 
of the University of Lausanne (Switzerland).}
\end{deluxetable}
\clearpage

%%%%%%%%%%%%%%%%%%%%%%%%%%%%%%% TABLE 4 %%%%%%%%%%%%%%%%%%%%%%%%%%%%%%%%%%

\begin{deluxetable}{lrrr} 
\tabletypesize{\footnotesize}
\tablewidth{0pc} 
\tablecaption{2MASS Photometry\label{tab2MASS}}
\tablehead{ 
\colhead{Star}        &  
\colhead{$J_{2M}$}    & 
\colhead{$H_{2M}$}    & 
\colhead{$K_{2M}$}    }
\startdata
    HD 886 & $ 3.500\pm0.264$ & $ 3.638\pm0.198$ & $ 3.770\pm0.288$ \\
   HD 9132 & $ 5.385\pm0.248$ & $ 5.026\pm0.026$ & $ 4.963\pm0.020$ \\
  HD 10250 & $ 5.267\pm0.286$ & $ 5.264\pm0.026$ & $ 5.215\pm0.017$ \\
  HD 21790 & $ 5.282\pm0.228$ & $ 4.960\pm0.036$ & $ 4.886\pm0.026$ \\
  HD 29646 & $ 5.682\pm0.019$ & $ 5.726\pm0.017$ & $ 5.705\pm0.021$ \\
  HD 32630 & $ 3.611\pm0.262$ & $ 3.761\pm0.238$ & $ 3.857\pm0.292$ \\
  HD 38899 & $ 5.024\pm0.290$ & $ 4.975\pm0.036$ & $ 4.983\pm0.016$ \\
  HD 45557 & $ 5.755\pm0.018$ & $ 5.803\pm0.027$ & $ 5.754\pm0.017$ \\
  HD 47105 & $ 1.729\pm0.244$ & $ 1.836\pm0.202$ & $ 1.917\pm0.220$ \\
  HD 58142 & $ 4.695\pm0.234$ & $ 4.693\pm0.033$ & $ 4.572\pm0.016$ \\
  HD 61831 & $ 5.218\pm0.037$ & $ 5.328\pm0.036$ & $ 5.338\pm0.017$ \\
  HD 66591 & $ 5.161\pm0.037$ & $ 5.259\pm0.034$ & $ 5.260\pm0.018$ \\
  HD 77002 & $ 5.326\pm0.043$ & $ 5.415\pm0.029$ & $ 5.413\pm0.021$ \\
  HD 79447 & $ 4.650\pm0.248$ & $ 4.565\pm0.270$ & $ 4.441\pm0.036$ \\
  HD 87901 & $ 1.665\pm0.314$ & $ 1.658\pm0.186$ & $ 1.640\pm0.212$ \\
  HD 93194 & $ 5.056\pm0.017$ & $ 5.109\pm0.029$ & $ 5.094\pm0.017$ \\
  HD 98664 & $ 4.366\pm0.302$ & $ 4.325\pm0.204$ & $ 4.139\pm0.036$ \\
 HD 103287 & $ 2.381\pm0.290$ & $ 2.487\pm0.174$ & $ 2.429\pm0.288$ \\
 HD 106911 & $ 4.687\pm0.206$ & $ 4.541\pm0.246$ & $ 4.556\pm0.026$ \\
 HD 115823 & $ 5.727\pm0.032$ & $ 5.760\pm0.045$ & $ 5.802\pm0.021$ \\
 HD 121263 & $ 3.022\pm0.254$ & $ 3.083\pm0.212$ & $ 3.220\pm0.250$ \\
 HD 121743 & $ 4.628\pm0.282$ & $ 4.461\pm0.264$ & $ 4.491\pm0.016$ \\
 HD 121790 & $ 4.721\pm0.214$ & $ 4.597\pm0.226$ & $ 4.469\pm0.036$ \\
 HD 125238 & $ 3.970\pm0.228$ & $ 3.893\pm0.210$ & $ 4.102\pm0.288$ \\
 HD 129116 & $ 4.637\pm0.244$ & $ 4.628\pm0.076$ & $ 4.487\pm0.026$ \\
 HD 147394 & $ 3.927\pm0.210$ & $ 4.085\pm0.236$ & $ 4.285\pm0.017$ \\
 HD 153808 & $ 3.554\pm0.174$ & $ 3.641\pm0.176$ & $ 3.916\pm0.016$ \\
 HD 158094 & $ 3.698\pm0.232$ & $ 3.651\pm0.224$ & $ 3.710\pm0.198$ \\
 HD 160762 & $ 4.267\pm0.270$ & $ 4.349\pm0.258$ & $ 4.228\pm0.016$ \\
 HD 172167 & $-0.177\pm0.206$ & $-0.029\pm0.146$ & $ 0.129\pm0.186$ \\
 HD 177724 & $ 3.084\pm0.330$ & $ 3.048\pm0.280$ & $ 2.876\pm0.360$ \\
 HD 182255 & $ 5.404\pm0.018$ & $ 5.480\pm0.020$ & $ 5.485\pm0.022$ \\
 HD 188665 & $ 5.397\pm0.017$ & $ 5.468\pm0.026$ & $ 5.516\pm0.034$ \\
 HD 192907 & $ 4.506\pm0.260$ & $ 4.418\pm0.036$ & $ 4.431\pm0.017$ \\
 HD 193432 & $ 4.909\pm0.218$ & $ 4.856\pm0.031$ & $ 4.807\pm0.016$ \\
 HD 196867 & $ 3.904\pm0.288$ & $ 3.890\pm0.236$ & $ 3.826\pm0.015$ \\
 HD 199081 & $ 5.253\pm0.260$ & $ 5.105\pm0.024$ & $ 5.103\pm0.016$ \\
 HD 201908 & $ 6.015\pm0.027$ & $ 6.108\pm0.045$ & $ 6.084\pm0.024$ \\
 HD 210419 & $ 6.234\pm0.024$ & $ 6.314\pm0.059$ & $ 6.211\pm0.016$ \\
 HD 214923 & $ 3.538\pm0.256$ & $ 3.527\pm0.216$ & $ 3.566\pm0.274$ \\
 HD 215573 & $ 5.668\pm0.044$ & $ 5.649\pm0.027$ & $ 5.650\pm0.020$ \\
 HD 218045 & $ 2.535\pm0.270$ & $ 2.744\pm0.220$ & $ 2.647\pm0.306$ \\
 HD 222439 & $ 4.624\pm0.264$ & $ 4.595\pm0.218$ & $ 4.571\pm0.354$ \\
 HD 222661 & $ 4.269\pm0.344$ & $ 4.622\pm0.018$ & $ 4.594\pm0.018$ \\
 HD 225132 & $ 4.151\pm0.222$ & $ 4.610\pm0.029$ & $ 4.564\pm0.017$ \\
\enddata
\tablecomments{The 2MASS data were collected from
the NASA/IPAC Infrared Science Archive, which is operated 
by the Jet Propulsion Laboratory, California 
Institute of Technology.}
\end{deluxetable}
\clearpage

%%%%%%%%%%%%%%%%%%%%%%%%%%%%%%% TABLE 5 %%%%%%%%%%%%%%%%%%%%%%%%%%%%%%%%%%

\begin{deluxetable}{lrrrcccc} 
\tabletypesize{\footnotesize}
\tablewidth{0pc} 
\tablecaption{Best-Fit Parameters for Calibration Stars\label{tabPARMS}}
\tablehead{ 
\colhead{Star}              &  
\colhead{$T_{eff}$}         & 
\colhead{[m/H]}             & 
\colhead{$v_{turb}$}        & 
\colhead{$R/R_{\sun}$}      & 
\colhead{$E(B-V)$}          & 
\colhead{$\log g(Spec)$\tablenotemark{a}} &
\colhead{$\log g(Newton)$\tablenotemark{b}}  \\
\colhead{}                  &  
\colhead{(K)}               & 
\colhead{}                  & 
\colhead{$\rm (km/s)$}      & 
\colhead{}                  & 
\colhead{(mag)}             & 
\colhead{}                  &
\colhead{}                  }
\startdata
   HD886 & $ 21179\pm 237$ & $ 0.07\pm 0.10$ & $  1.3\pm  0.9$ & $ 4.80\pm0.39$ & $0.000$ & $ 3.98\pm0.06$ & $ 3.98$ \\
  HD9132 & $  9198\pm  46\phn$ & $-0.21\pm 0.09$ & $  2.0\pm  0.8$ & $ 2.39\pm0.13$ & $0.000$ & $ 4.04\pm0.04$ & $ 4.04$ \\
 HD10250 & $ 10141\pm  61\phn$ & $ 0.03\pm 0.09$ & $  0.3\pm  0.8$ & $ 2.63\pm0.13$ & $0.000$ & $ 3.98\pm0.04$ & $ 4.02$ \\
 HD21790 & $ 11548\pm  74\phn$ & $-0.64\pm 0.09$ & $  1.6\pm  0.7$ & $ 3.91\pm0.34$ & $0.003\pm0.005$ & $ 3.78\pm0.06$ & $ 3.80$ \\
 HD29646 & $  9693\pm  68\phn$ & $ 0.22\pm 0.10$ & $  0.0\pm  0.3$ & $ 2.53\pm0.24$ & $0.000$ & $ 3.99\pm0.07$ & $ 4.02$ \\
 HD32630 & $ 17201\pm 173$ & $-0.07\pm 0.07$ & $  0.0\pm  0.1$ & $ 3.25\pm0.18$ & $0.000$ & $ 4.13\pm0.04$ & $ 4.15$ \\
 HD38899 & $ 11089\pm  67\phn$ & $-0.25\pm 0.10$ & $  1.5\pm  0.8$ & $ 2.66\pm0.18$ & $0.000$ & $ 4.05\pm0.05$ & $ 4.06$ \\
 HD45557 & $  9595\pm  39\phn$ & $ 0.05\pm 0.05$ & $  0.2\pm  0.5$ & $ 2.15\pm0.08$ & $0.000$ & $ 4.13\pm0.03$ & $ 4.90$ \\
 HD47105 & $  9113\pm  89\phn$ & $ 0.04\pm 0.14$ & $  0.0\pm  0.5$ & $ 4.95\pm0.32$ & $0.000$ & $ 3.53\pm0.04$ & $ 3.53$ \\
 HD58142 & $  9416\pm  47\phn$ & $-0.24\pm 0.11$ & $  1.1\pm  0.8$ & $ 3.30\pm0.19$ & $0.000$ & $ 3.83\pm0.04$ & $ 3.83$ \\
 HD61831 & $ 17917\pm 293$ & $-0.44\pm 0.42$ & $  0.0\pm  0.1$ & $ 3.86\pm0.34$ & $0.023\pm0.009$ & $ 3.99\pm0.07$ & $ 4.04$ \\
 HD66591 & $ 17162\pm 281$ & $-0.11\pm 0.17$ & $  0.0\pm  0.2$ & $ 3.77\pm0.27$ & $0.011\pm0.007$ & $ 4.03\pm0.05$ & $ 4.03$ \\
 HD77002 & $ 18773\pm 223$ & $ 0.07\pm 0.13$ & $  0.9\pm  0.7$ & $ 3.82\pm0.37$ & $0.000$ & $ 4.08\pm0.08$ & $ 4.08$ \\
 HD79447 & $ 17365\pm 290$ & $ 0.12\pm 0.21$ & $  0.0\pm  0.5$ & $ 5.03\pm0.35$ & $0.019\pm0.007$ & $ 3.83\pm0.05$ & $ 3.83$ \\
 HD87901 & $ 12194\pm  62\phn$ & $-0.37\pm 0.05$ & $  0.1\pm  0.7$ & $ 3.57\pm0.09$ & $0.000$ & $ 3.54\pm0.09$ & $ 3.89$ \\
 HD93194 & $ 15690\pm 247$ & $-0.33\pm 0.16$ & $  0.0\pm  0.0$ & $ 3.63\pm0.27$ & $0.010\pm0.007$ & $ 3.83\pm0.09$ & $ 4.01$ \\
 HD98664 & $ 10220\pm  52\phn$ & $-0.36\pm 0.08$ & $  1.0\pm  0.7$ & $ 3.38\pm0.16$ & $0.000$ & $ 3.83\pm0.03$ & $ 3.85$ \\
HD103287 & $  9336\pm  52\phn$ & $-0.19\pm 0.08$ & $  1.8\pm  0.7$ & $ 3.04\pm0.08$ & $0.000$ & $ 3.79\pm0.03$ & $ 3.88$ \\
HD106911 & $ 14495\pm 157$ & $-0.40\pm 0.13$ & $  0.5\pm  0.7$ & $ 2.84\pm0.13$ & $0.002\pm0.004$ & $ 4.03\pm0.05$ & $ 4.15$ \\
HD115823 & $ 14244\pm 173$ & $-0.13\pm 0.15$ & $  0.0\pm  0.0$ & $ 2.38\pm0.11$ & $0.000$ & $ 4.26\pm0.04$ & $ 4.27$ \\
HD121263 & $ 23561\pm 283$ & $-0.24\pm 0.10$ & $  2.2\pm  0.6$ & $ 5.80\pm0.53$ & $0.000$ & $ 3.84\pm0.08$ & $ 3.91$ \\
HD121743 & $ 21638\pm 388$ & $ 0.03\pm 0.11$ & $  1.2\pm  1.0$ & $ 4.19\pm0.35$ & $0.006\pm0.006$ & $ 4.08\pm0.07$ & $ 4.10$ \\
HD121790 & $ 21411\pm 377$ & $-0.05\pm 0.12$ & $  1.2\pm  1.0$ & $ 3.74\pm0.34$ & $0.011\pm0.007$ & $ 4.15\pm0.07$ & $ 4.18$ \\
HD125238 & $ 18605\pm 221$ & $ 0.19\pm 0.15$ & $  0.0\pm  0.5$ & $ 4.05\pm0.33$ & $0.000$ & $ 3.94\pm0.07$ & $ 4.03$ \\
HD129116 & $ 18445\pm 344$ & $-0.12\pm 0.13$ & $  0.0\pm  0.1$ & $ 2.93\pm0.12$ & $0.010\pm0.008$ & $ 4.23\pm0.03$ & $ 4.27$ \\
HD147394 & $ 15615\pm 301$ & $ 0.24\pm 0.18$ & $  0.0\pm  0.5$ & $ 3.55\pm0.19$ & $0.012\pm0.007$ & $ 4.02\pm0.05$ & $ 4.03$ \\
HD153808 & $ 10197\pm  57\phn$ & $-0.13\pm 0.10$ & $  0.7\pm  0.8$ & $ 2.72\pm0.07$ & $0.000$ & $ 3.98\pm0.02$ & $ 4.00$ \\
HD158094 & $ 11962\pm  86\phn$ & $-0.62\pm 0.10$ & $  1.9\pm  0.7$ & $ 3.12\pm0.15$ & $0.000$ & $ 3.81\pm0.06$ & $ 3.98$ \\
HD160762 & $ 18070\pm 294$ & $-0.22\pm 0.22$ & $  0.0\pm  0.0$ & $ 5.29\pm0.45$ & $0.032\pm0.007$ & $ 3.82\pm0.06$ & $ 3.82$ \\
HD172167 & $  9549\pm  41\phn$ & $-0.51\pm 0.07$ & $  0.0\pm  0.4$ & $ 2.75\pm0.03$ & $0.000$ & $ 3.96\pm0.01$ & $ 3.96$ \\
HD177724 & $  9308\pm  59\phn$ & $-0.40\pm 0.16$ & $  0.0\pm  0.1$ & $ 2.38\pm0.06$ & $0.000$ & $ 3.60\pm0.14$ & $ 4.05$ \\
HD182255 & $ 14650\pm 205$ & $-0.31\pm 0.18$ & $  0.8\pm  1.1$ & $ 2.70\pm0.19$ & $0.012\pm0.008$ & $ 4.19\pm0.05$ & $ 4.19$ \\
HD188665 & $ 14893\pm 214$ & $-0.17\pm 0.16$ & $  0.8\pm  0.8$ & $ 4.30\pm0.45$ & $0.008\pm0.006$ & $ 3.82\pm0.08$ & $ 3.86$ \\
HD192907 & $ 10174\pm  55\phn$ & $-0.32\pm 0.09$ & $  0.7\pm  0.7$ & $ 4.45\pm0.22$ & $0.000$ & $ 3.66\pm0.03$ & $ 3.66$ \\
HD193432 & $ 10213\pm  68\phn$ & $-0.08\pm 0.11$ & $  0.2\pm  0.7$ & $ 3.09\pm0.26$ & $0.000$ & $ 3.91\pm0.06$ & $ 3.91$ \\
HD196867 & $ 11243\pm  85\phn$ & $-0.49\pm 0.09$ & $  1.8\pm  0.7$ & $ 3.92\pm0.20$ & $0.022\pm0.007$ & $ 3.72\pm0.04$ & $ 3.79$ \\
HD199081 & $ 15484\pm 250$ & $-0.16\pm 0.16$ & $  0.0\pm  0.0$ & $ 3.87\pm0.36$ & $0.015\pm0.008$ & $ 3.95\pm0.07$ & $ 3.96$ \\
HD201908 & $ 11574\pm 58\phn$ & $-0.56\pm 0.06$ & $  1.7\pm  0.5$ & $ 2.45\pm0.13$ & $0.000$ & $ 4.03\pm0.05$ & $ 4.13$ \\
HD210419 & $  9499\pm 75\phn$ & $ 0.07\pm 0.15$ & $  0.0\pm  0.4$ & $ 2.21\pm0.20$ & $0.000$ & $ 3.60\pm0.18$ & $ 4.11$ \\
HD214923 & $ 11190\pm 55\phn$ & $-0.38\pm 0.07$ & $  1.6\pm  0.6$ & $ 4.03\pm0.22$ & $0.000$ & $ 3.67\pm0.05$ & $ 3.77$ \\
HD215573 & $ 14347\pm 138$ & $-0.31\pm 0.16$ & $  0.8\pm  0.9$ & $ 2.84\pm0.20$ & $0.000$ & $ 4.14\pm0.05$ & $ 4.14$ \\
HD218045 & $  9765\pm 63\phn$ & $-0.02\pm 0.10$ & $  0.2\pm  0.7$ & $ 4.72\pm0.14$ & $0.000$ & $ 3.51\pm0.03$ & $ 3.60$ \\
HD222439 & $ 11361\pm 66\phn$ & $-0.36\pm 0.09$ & $  1.6\pm  0.8$ & $ 2.31\pm0.09$ & $0.000$ & $ 4.10\pm0.03$ & $ 4.17$ \\
HD222661 & $ 10504\pm 91\phn$ & $-0.35\pm 0.14$ & $  1.5\pm  1.0$ & $ 1.94\pm0.06$ & $0.000$ & $ 4.22\pm0.03$ & $ 4.26$ \\
HD225132 & $ 10291\pm 108$ & $-0.15\pm 0.11$ & $  0.0\pm  0.2$ & $ 2.82\pm0.16$ & $0.000$ & $ 3.88\pm0.05$ & $ 3.98$ \\
\tablenotetext{a}{The quantity $\log g(Spec)$ refers to the ``Spectroscopic'' surface gravity, which is used to characterize the best-fitting \atlas\/ model atmospheres. $\log g(Spec)$ is based on the Newtonian surface gravity, but reduced to simulate the lower atmospheric pressures resulting from rotationally-induced centrifugal force.  See Eq. \ref{gspec} in \S 2.}
\tablenotetext{b}{The quantity $\log g(Newton)$ refers to the ``Newtonian'' surface gravity, corresponding to $GM/R^2$, and is determined as described in \S 2.}
\enddata
\end{deluxetable}
\clearpage

%%%%%%%%%%%%%%%%%%%%%%%%%%%%%%% TABLE 6 %%%%%%%%%%%%%%%%%%%%%%%%%%%%%%%%%%

\begin{deluxetable}{ll}
\tablewidth{0pc}
\tablecaption{Sources for Filter Sensitivities Curves
\label{tabFILT}}
\tablehead{
\colhead{Filter}                   & 
\colhead{ Source}}
\startdata
$UBV$    & Buser \& Kurucz (1978)    \\
$RIJK$   & Johnson (1965)            \\
$H$      & Bessell \& Brett (1988)   \\
$uvby$   & Matsushima (1969)         \\ 
$Geneva$ &  Nicolet (2004)           \\
$2MASS$  & Cutri et~al. (2000)       \\
\enddata
\end{deluxetable}
\clearpage

%%%%%%%%%%%%%%%%%%%%%%%%%%%%%%% TABLE 7 %%%%%%%%%%%%%%%%%%%%%%%%%%%%%%%%%%

\begin{deluxetable}{ccccc}
\tablewidth{0pc}
\tabletypesize{\footnotesize}
\tablecaption{Uncertainties in Photometric Transformations
\label{tabSCAT}}
\tablehead{
\colhead{Photometric}                  & 
\colhead{Actual}                       &
\colhead{Random Errors}                &
\colhead{Implied Mean}                 &
\colhead{Expected}                      \\

\colhead{Index}                        &
\colhead{Transformation}               &
\colhead{in Synthetic}                 &
\colhead{Observational}                &
\colhead{Observational}               \\

\colhead{  }                           & 
\colhead{Scatter\tablenotemark{a}}     &
\colhead{Photometry\tablenotemark{b}}  &
\colhead{Error\tablenotemark{c}}       &

\colhead{Error\tablenotemark{d}}      \\
\colhead{  }                           & 
\colhead{(mag)}                        &
\colhead{(mag)}                        &
\colhead{(mag)}                        &
\colhead{(mag)}                         }
\startdata
\multicolumn{5}{l}{Johnson Photometry} \\
$U-B$    &  0.021  & 0.010 & 0.018   & 0.016    \\
$B-V$    &  0.011  & 0.005 & 0.010   & 0.011    \\
$V-R$    &  0.026  & 0.003 & 0.026   & \nodata  \\
$R-I$    &  0.021  & 0.004 & 0.021   & \nodata  \\
$J$      &  0.013  & 0.021 & \nodata & \nodata  \\
$H$      &  0.033  & 0.022 & 0.025   & \nodata  \\
$K$      &  0.019  & 0.023 & \nodata & \nodata  \\
\multicolumn{5}{l}{Str\"{o}mgren Photometry}        \\
$b-y$    &  0.006  & 0.003 & 0.005   & 0.008    \\
$m1$     &  0.008  & 0.003 & 0.007   & 0.011    \\ 
$c1$     &  0.019  & 0.014 & 0.013   & 0.012    \\
$\beta$  &  0.012  & 0.008 & 0.009   & 0.010    \\
\multicolumn{5}{l}{Geneva Photometry}           \\
$U-B$    & 0.021   & 0.014 & 0.016   & 0.011    \\
$V-B$    & 0.013   & 0.006 & 0.012   & 0.008    \\
$B1-B$   & 0.008   & 0.003 & 0.007   & 0.008    \\
$B2-B$   & 0.007   & 0.003 & 0.006   & 0.008    \\
$V1-B$   & 0.012   & 0.006 & 0.010   & 0.008    \\
$G-B$    & 0.014   & 0.007 & 0.012   & 0.009    \\
\multicolumn{5}{l}{{\it 2MASS} Photometry}      \\
$J_{2M}$ &  0.037  & 0.021 & 0.030   & 0.029    \\
$H_{2M}$ &  0.058  & 0.022 & 0.054   & 0.036    \\
$K_{2M}$ &  0.036  & 0.023 & 0.028   & 0.022    \\
\enddata
\tablenotetext{a}{Standard deviation of the observed photometry compared 
with the transformation relations shown in Figures 4 and 5.}
\tablenotetext{b}{For each star, we computed the standard deviations about 
the mean of the 15 photometric indices from the 100 Monte Carlo SED fits.  
The values listed are the RMS means (averaged over all the stars) of those 
standard deviations.}  
\tablenotetext{c}{Mean value of the random photometric errors, computed 
assuming that the scatter seen in the transformation relations arises from 
the quadratic sum of the random error in the synthetic photometry and 
random observational error in the data.} 
\tablenotetext{d}{Expected errors for Johnson $U-B$ and $B-V$ indices
are from the statistical analysis of FitzGerald 1973.  Those for the
Str\"{o}mgren system are from Lindemann 1974.  Those for the Geneva are
derived from the analysis of Rufener 1981. For the \tmass\/ magnitudes,
the errors listed are the RMS values derived from the actual
observational errors for the stars in our sample.  Only the
well-determined magnitudes (those with errors less than 0.1 mag) were
used in this calculation (see Figure \ref{fig2MASS}).}
\end{deluxetable}
\clearpage

%%%%%%%%%%%%%%%%%%%%%%%%%%%%%%% TABLE 8a %%%%%%%%%%%%%%%%%%%%%%%%%%%%%%%%%%

\begin{deluxetable}{rrrrrrrrrrrr} 
\tabletypesize{\scriptsize}
\tablewidth{0pt}
\tablenum{8a}
\tablecaption{Synthetic Johnson Photometry for ATLAS9 [m/H] =-1.5, 
v$_{turb}$ = 0 km/s Models [The complete version of this table is in the 
electronic edition of the Journal.  The printed edition contains only a sample.]}
\tablehead{ 
\colhead{$T_{eff}$}      & 
\colhead{$\log g$}       & 
\colhead{[m/H]}          & 
\colhead{$v_{turb}$}     & 
\colhead{$V$}            & 
\colhead{$U-B$}          & 
\colhead{$B-V$}          & 
\colhead{$V-R$}          & 
\colhead{$R-I$}          & 
\colhead{$J$}            & 
\colhead{$H$}            & 
\colhead{$K$}} 
\startdata
  9000. &   1.40 & -1.5 &  0.0 & -40.234 &  -0.199 &   0.008 &   0.083 &   0.045 & -40.336 & -40.371 & -40.372 \\
  9000. &   1.50 & -1.5 &  0.0 & -40.243 &  -0.176 &  -0.002 &   0.074 &   0.044 & -40.339 & -40.372 & -40.372 \\
  9000. &   2.00 & -1.5 &  0.0 & -40.268 &  -0.090 &  -0.031 &   0.048 &   0.038 & -40.346 & -40.371 & -40.373 \\
  9000. &   2.50 & -1.5 &  0.0 & -40.278 &  -0.026 &  -0.033 &   0.040 &   0.029 & -40.351 & -40.371 & -40.376 \\
  9000. &   3.00 & -1.5 &  0.0 & -40.281 &   0.024 &  -0.017 &   0.039 &   0.019 & -40.355 & -40.371 & -40.380 \\
\enddata
\end{deluxetable}
\clearpage

%%%%%%%%%%%%%%%%%%%%%%%%%%%%%%% TABLE 9a %%%%%%%%%%%%%%%%%%%%%%%%%%%%%%%%%%

\begin{deluxetable}{rrrrrrrr} 
\tabletypesize{\scriptsize}
\tablewidth{0pt}
\tablenum{9a}
\tablecaption{Synthetic Stromgren Photometry for ATLAS9 [m/H] =-1.5, v$_{turb}$ = 0 km/s Models [The complete version of this table is in the electronic edition of the Journal.  The printed edition contains only a sample.]}
\tablehead{ 
\colhead{$T_{eff}$}      & 
\colhead{$\log g$}       & 
\colhead{[m/H]}          & 
\colhead{$v_{turb}$}     & 
\colhead{$b-y$}          & 
\colhead{$m1$}           & 
\colhead{$c1$}           & 
\colhead{$\beta$}        } 
\startdata
  9000. &   1.40 & -1.5 &  0.0 &   0.056 &   0.063 &   1.133 &   2.627 \\
  9000. &   1.50 & -1.5 &  0.0 &   0.048 &   0.066 &   1.184 &   2.644 \\
  9000. &   2.00 & -1.5 &  0.0 &   0.020 &   0.080 &   1.326 &   2.724 \\
  9000. &   2.50 & -1.5 &  0.0 &   0.010 &   0.096 &   1.352 &   2.794 \\
  9000. &   3.00 & -1.5 &  0.0 &   0.009 &   0.117 &   1.310 &   2.848 \\
\enddata
\end{deluxetable}
\clearpage

%%%%%%%%%%%%%%%%%%%%%%%%%%%%%%% TABLE 10a %%%%%%%%%%%%%%%%%%%%%%%%%%%%%%%%%%

\newpage
\begin{deluxetable}{rrrrrrrrrrrr} 
\tabletypesize{\scriptsize}
\tablewidth{0pt}
\tablenum{10a}
\tablecaption{Synthetic Geneva Photometry for ATLAS9 [m/H] =-1.5, v$_{turb}$ = 0 km/s Models [The complete version of this table is in the electronic edition of the Journal.  The printed edition contains only a sample.]}
\tablehead{ 
\colhead{$T_{eff}$}      & 
\colhead{$\log g$}       & 
\colhead{[m/H]}          & 
\colhead{$v_{turb}$}     & 
\colhead{$U-B$}          & 
\colhead{$V-B$}           & 
\colhead{$B1-B$}          & 
\colhead{$B2-B$}          & 
\colhead{$V1-B$}          & 
\colhead{$G-B$}           } 
\startdata
  9000. &   1.40 & -1.5 &  0.0 &   1.477 &   0.977 &   0.826 &   1.556 &   1.676 &   2.157 \\
  9000. &   1.50 & -1.5 &  0.0 &   1.520 &   0.987 &   0.826 &   1.557 &   1.686 &   2.170 \\
  9000. &   2.00 & -1.5 &  0.0 &   1.648 &   1.012 &   0.830 &   1.554 &   1.710 &   2.205 \\
  9000. &   2.50 & -1.5 &  0.0 &   1.688 &   1.007 &   0.843 &   1.544 &   1.705 &   2.202 \\
  9000. &   3.00 & -1.5 &  0.0 &   1.677 &   0.983 &   0.860 &   1.527 &   1.681 &   2.179 \\
\enddata
\end{deluxetable}
\clearpage

%%%%%%%%%%%%%%%%%%%%%%%%%%%%%%% TABLE 11a %%%%%%%%%%%%%%%%%%%%%%%%%%%%%%%%%%

\begin{deluxetable}{rrrrrrrrrrrr} 
\tabletypesize{\scriptsize}
\tablewidth{0pt}
\tablenum{11a}
\tablecaption{Synthetic 2MASS Photometry for ATLAS9 [m/H] =-1.5, v$_{turb}$ = 0 km/s Models [The complete version of this table is in the electronic edition of the Journal.  The printed edition contains only a sample.]}
\tablehead{ 
\colhead{$T_{eff}$}      & 
\colhead{$\log g$}       & 
\colhead{[m/H]}          & 
\colhead{$v_{turb}$}     & 
\colhead{$J_{2M}$}       & 
\colhead{$H_{2M}$}       & 
\colhead{$K_{2M}$}        } 
\startdata
  9000. &   1.40 & -1.5 &  0.0 & -40.381 & -40.383 & -40.425 \\
  9000. &   1.50 & -1.5 &  0.0 & -40.384 & -40.384 & -40.425 \\
  9000. &   2.00 & -1.5 &  0.0 & -40.390 & -40.384 & -40.426 \\
  9000. &   2.50 & -1.5 &  0.0 & -40.396 & -40.383 & -40.429 \\
  9000. &   3.00 & -1.5 &  0.0 & -40.401 & -40.383 & -40.434 \\
\enddata
\end{deluxetable}

%------------------------------------------------------------------------%
% FIGURES                                                                %
%------------------------------------------------------------------------%
%%%%%%%%%%%%%%%%%%%%%%%%%%%%%%% FIGURE 1 %%%%%%%%%%%%%%%%%%%%%%%%%%%%%%%%%%

%%%%%%%%%%%%%%%%%%%%%%%%%%%%%%% FIGURE 2 %%%%%%%%%%%%%%%%%%%%%%%%%%%%%%%%%%

\begin{figure}[ht]
\figurenum{2}
\epsscale{0.75}
\plotone{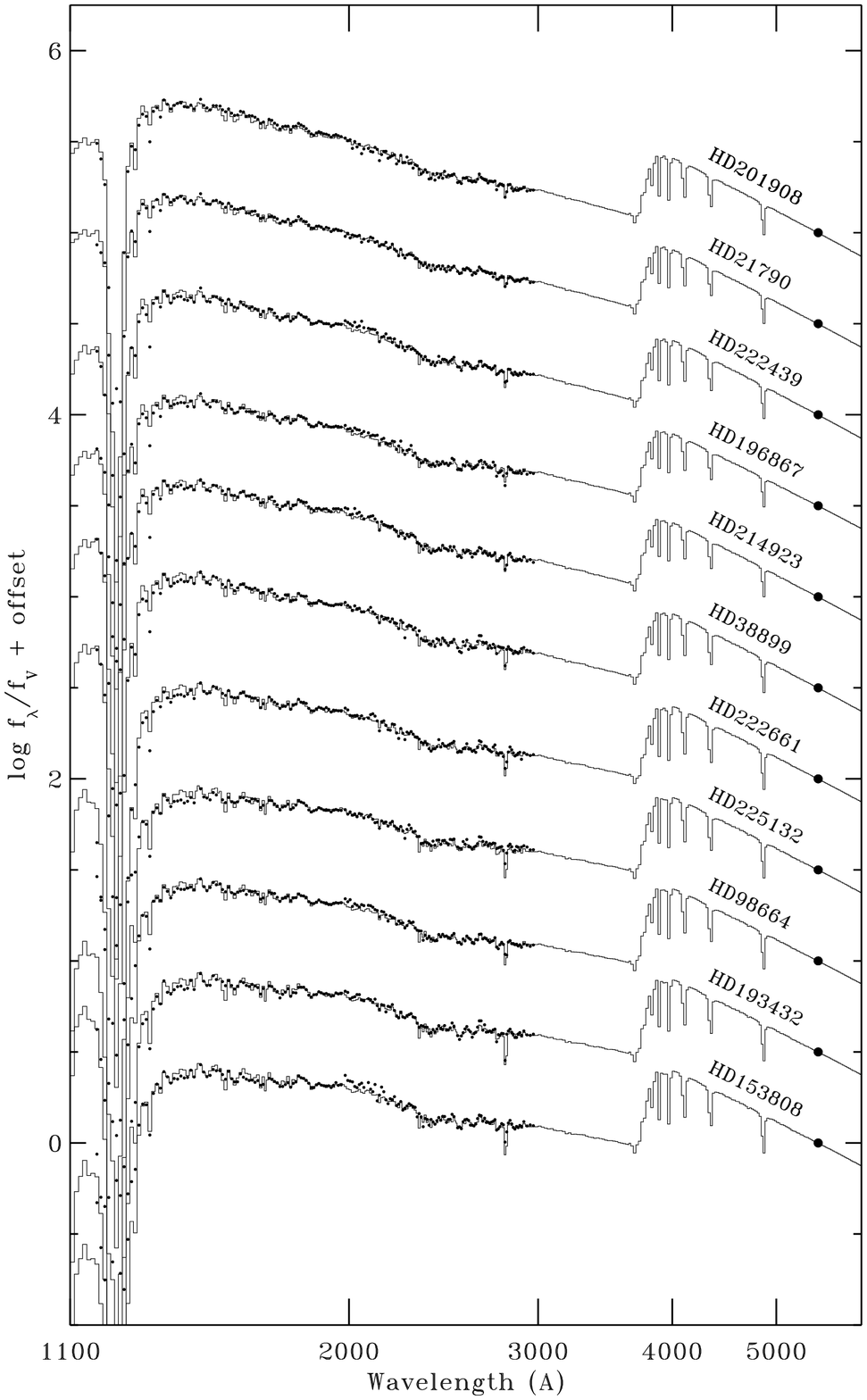}
\caption{Figure 2 continued
\label{figSEDb}}
\end{figure}

\begin{figure}[ht]
\figurenum{2}
\epsscale{0.75}
\plotone{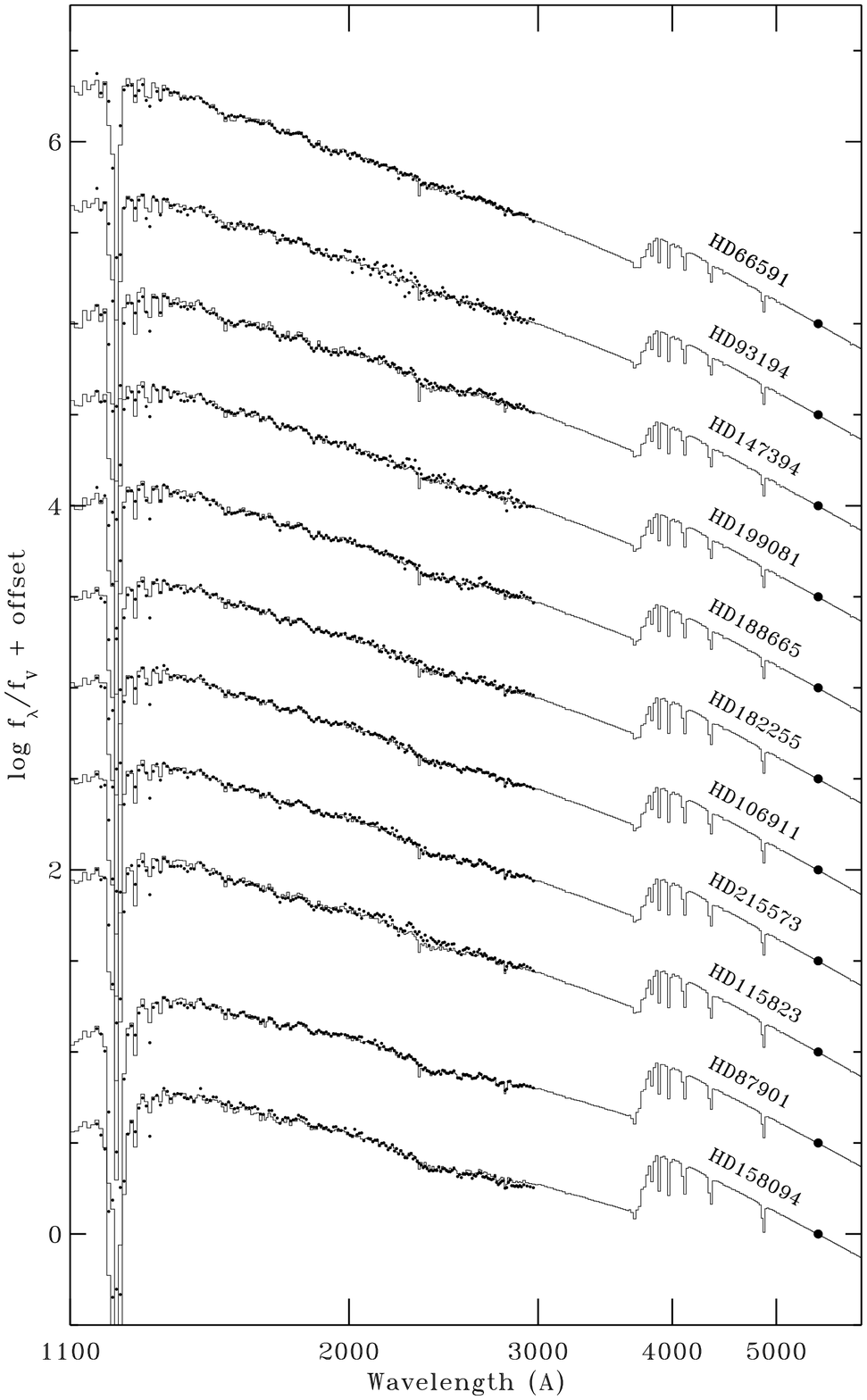}
\caption{Figure 2 continued
\label{figSEDc}}
\end{figure}

\begin{figure}[ht]
\figurenum{2}
\epsscale{0.75}
\plotone{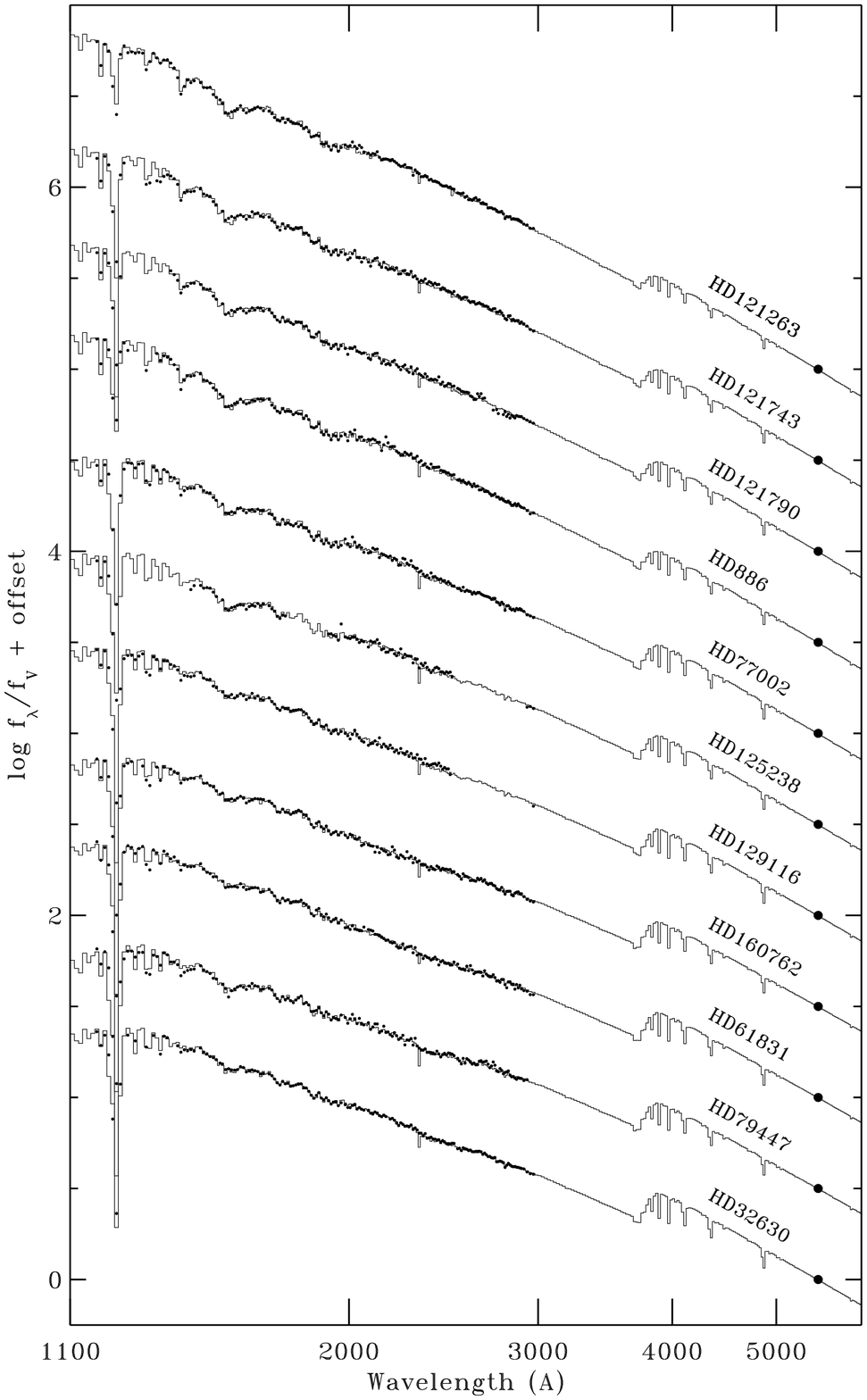}
\caption{Figure 2 continued
\label{figSEDd}}
\end{figure}

%%%%%%%%%%%%%%%%%%%%%%%%%%%%%%% FIGURE 3 %%%%%%%%%%%%%%%%%%%%%%%%%%%%%%%%%%

\begin{figure}[ht]
\figurenum{3}
\epsscale{1.0}
\plotone{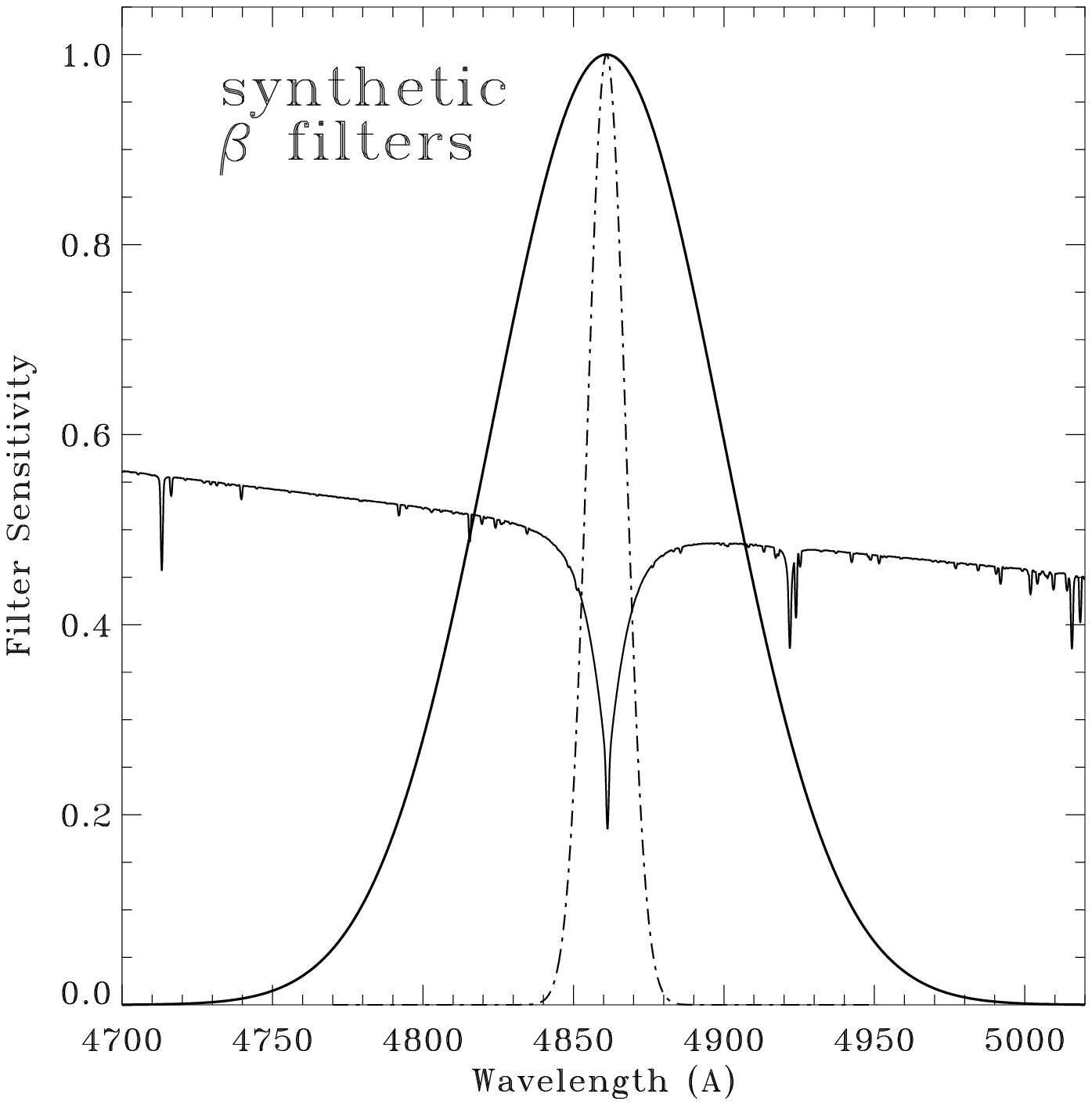}
\caption{The adopted models for the $\beta$ filter profiles, compared
with a synthetic spectrum for a typical B star (\teff\ = 15000 K, $\log
g = 4$, [m/H] = 0.0, \vturb\/ = 2~\kms).  The intermediate (solid
curve) and narrow (dash-dot curve) $\beta$ filters are represented by
Gaussians with FWHM values of 90 \AA\/ and 15 \AA, respectively.  The
$\beta$ index is the difference between the magnitudes measured by the
two filters.  We perform the synthetic photometry with the filters
scaled to maximum values of 1.0, as shown in the figure.
\label{figBETA}}
\end{figure}

%%%%%%%%%%%%%%%%%%%%%%%%%%%%%%% FIGURE 4 %%%%%%%%%%%%%%%%%%%%%%%%%%%%%%%%%%

\begin{figure}[ht]
\figurenum{4}
\epsscale{0.75}
\plotone{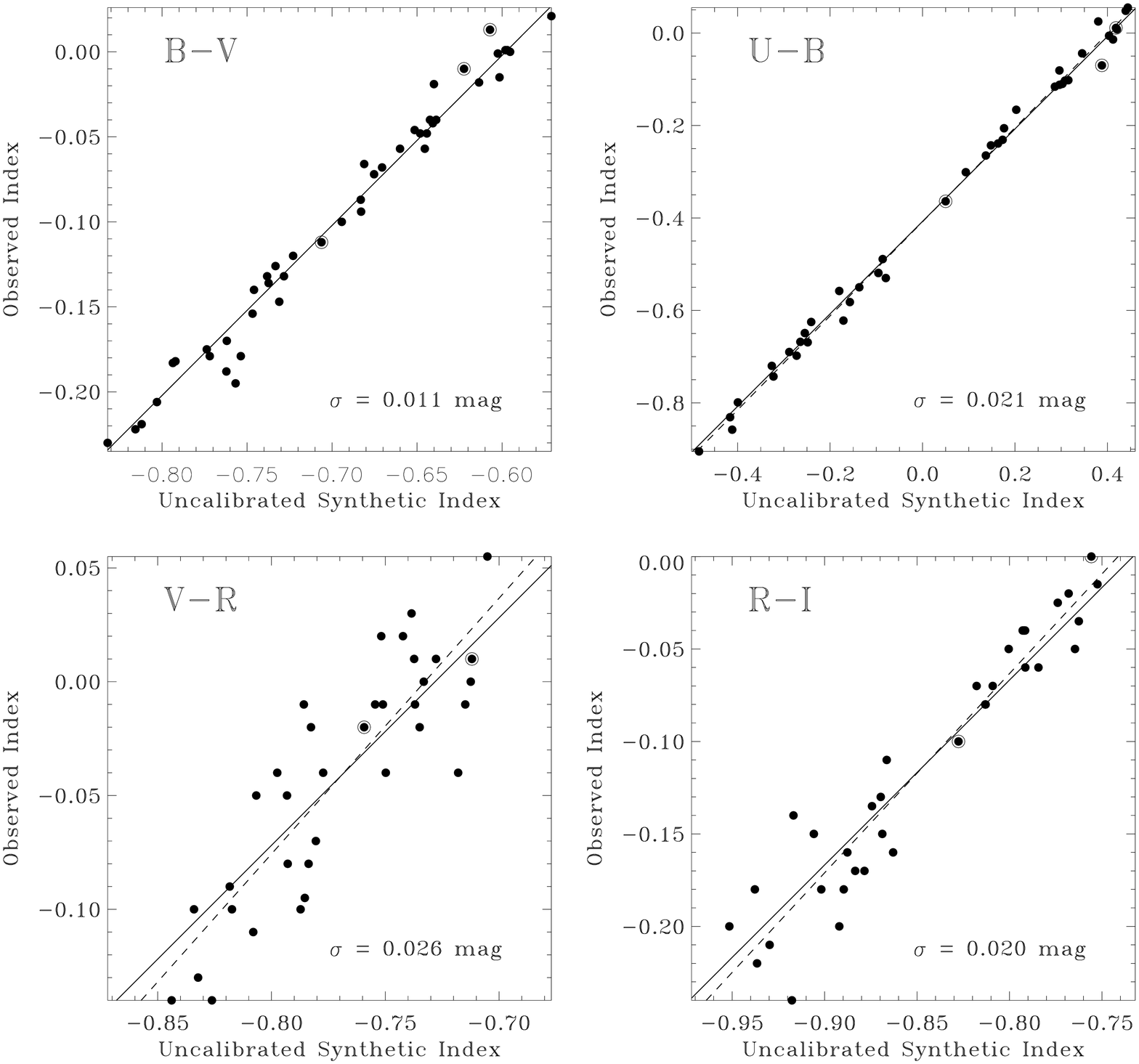}
\caption{Filter calibration relations for Johnson photometric indices.
Each panel shows the relationship between the uncalibrated synthetic
indices and the observed indices for the 45 calibration stars.  In all
the panels, the solid lines have a slope of unity and show the simple
mean offset between the observed and synthetic photometry.  This
provides an adequate calibration for $B-V$, but the other indices
require linear calibrations, as indicated by the dashed lines.  The
adopted relationships between the observed and synthetic indices are as
follows:  $B-V =  0.598 + (B-V)_{syn}$; $U-B= -0.408+1.018(U-B)_{syn}$;
$V-R = 0.823+1.124(V-R)_{syn}$; and $R-I =0.799+1.078(R-I)_{syn}$. Circled points show the positions of the three most rapid rotators in the sample: HD~87901, HD~177724, and HD~210419.  These are discussed in \S 7.2.
\label{figUBVRI}}
\end{figure}

%%%%%%%%%%%%%%%%%%%%%%%%%%%%%%% FIGURE 5 %%%%%%%%%%%%%%%%%%%%%%%%%%%%%%%%%%

\begin{figure}[ht]
\figurenum{5}
\epsscale{0.75}
\plotone{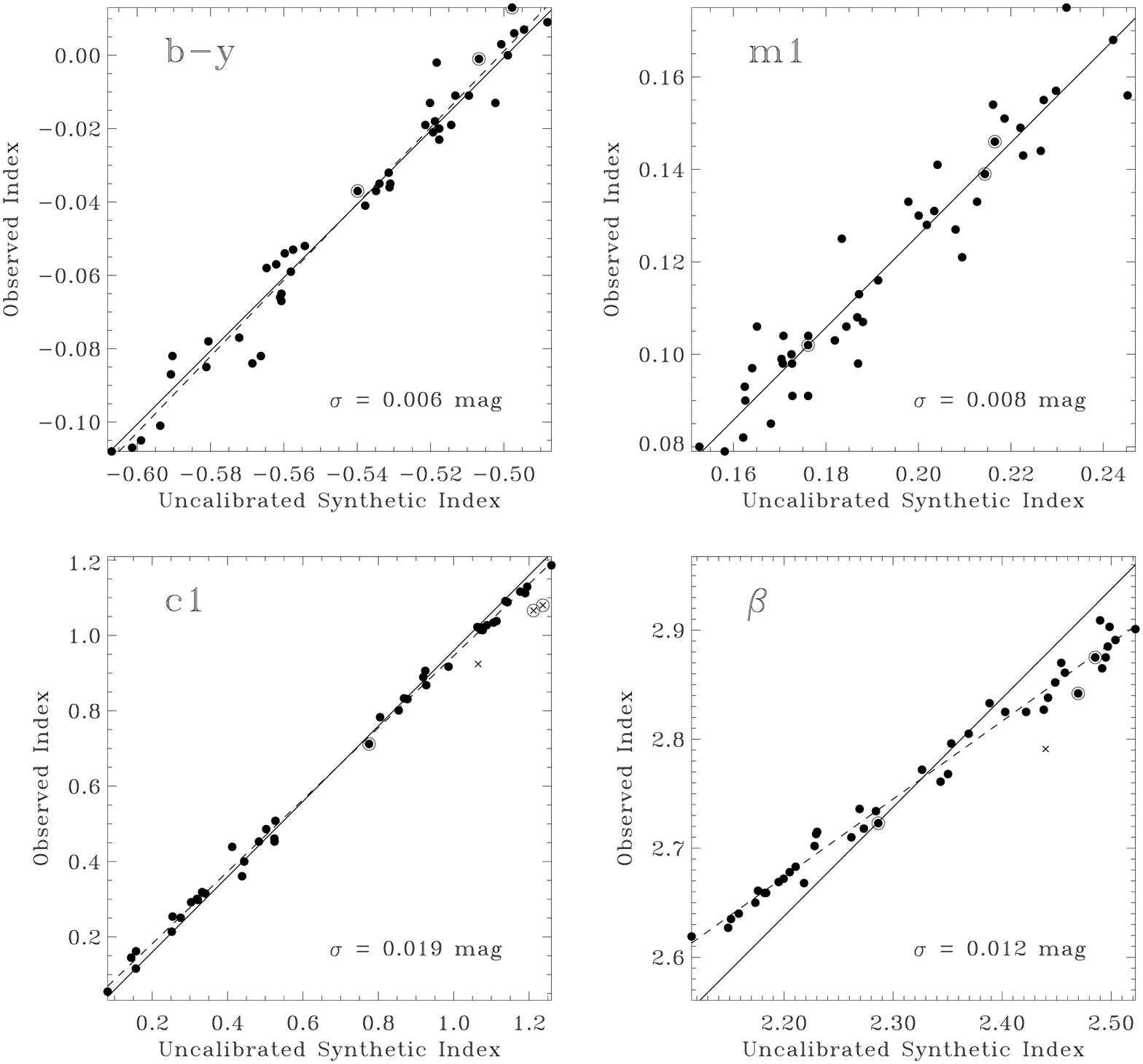} \caption{Same as Figure
\ref{figUBVRI}, but for the Str\"{o}mgren photometric indices. For
$m1$, a simple offset of unity slope provides an adequate calibration
(solid line) and is given by $m1 =  -0.074 + m1_{syn}$.  For $b-y$,
$c1$, and $\beta$, linear calibrations are adopted (dashed lines),
given by: $b-y = 0.521+1.040(b-y)_{syn}$;  $c1 = -0.009+0.955c1_{syn}$;
and $\beta = 1.101+0.715\beta_{syn}$.  In the $c1$ and $\beta$ panels,
the crosses indicate data points excluded from the calibrations, and
correspond to the stars HD 158303, HD 177724, and HD 210419 for $c1$
and HD 225132 for $\beta$.  As in Figure \ref{figUBVRI}, circled points
show the positions of the three most rapid rotators in the sample:
HD~87901, HD~177724, and HD~210419.
\label{figUVBY}}
\end{figure}

%%%%%%%%%%%%%%%%%%%%%%%%%%%%%%% FIGURE 6 %%%%%%%%%%%%%%%%%%%%%%%%%%%%%%%%%%

\begin{figure}[ht]
\figurenum{6}
\epsscale{0.65}
\plotone{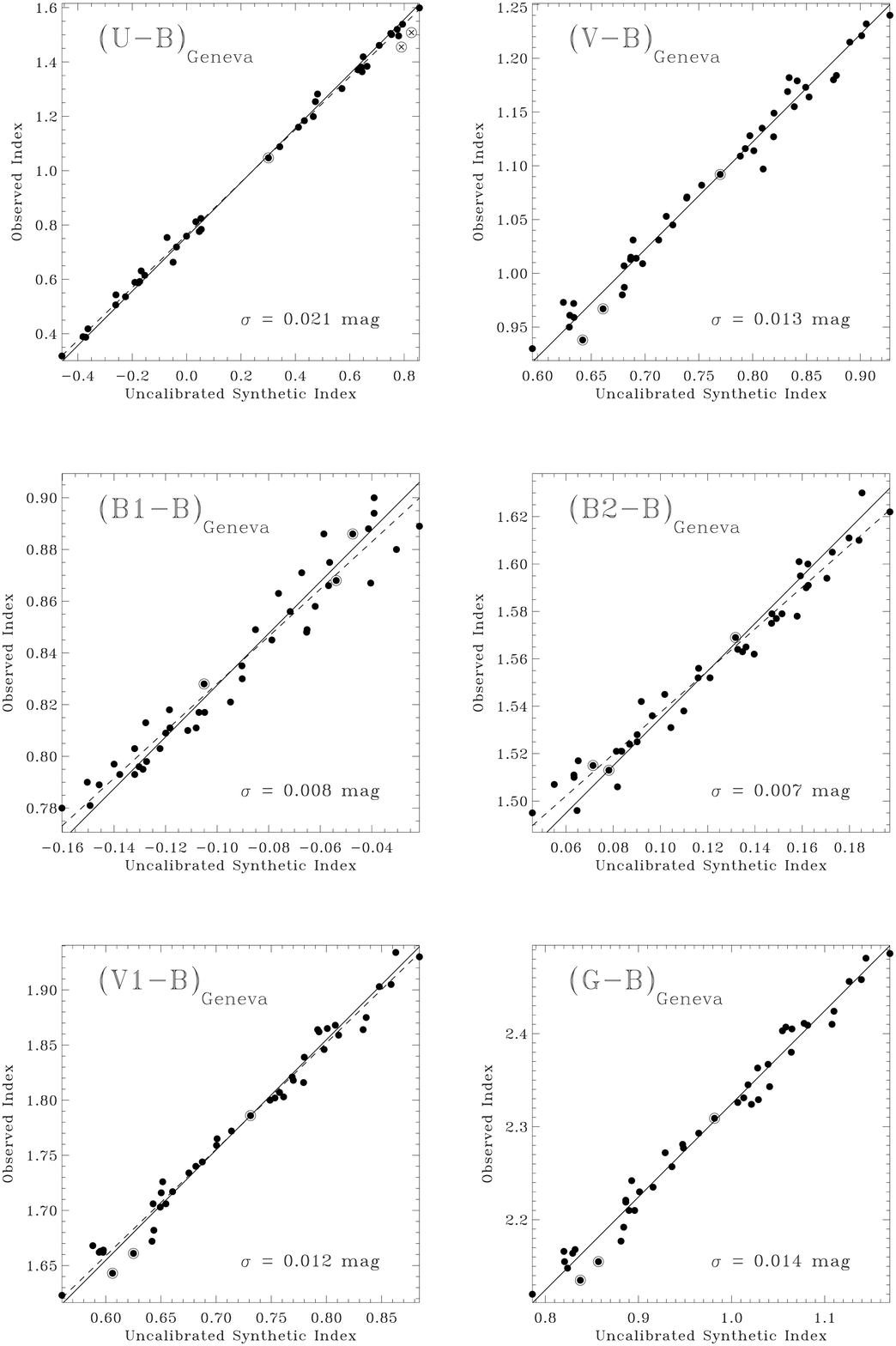}
\caption{Same as Figure \ref{figUBVRI}, but for the Geneva optical
colors. For $V-B$ and $G-B$, simple offsets of unity slope provide
adequate calibrations (solid lines) and are given by $V-B = 0.322 +
(V-B)_{syn}$ and $G-B = 1.324 + (U-B)_{syn}$.  For the other indices,
linear calibrations are adopted (dashed lines) and are given by $U-B =
0.764 + 0.969(U-B)_{syn}$; $B1-B = 0.920 + 0.914(B1-B)_{syn}$; $B2-B =
1.449 + 0.881(B2-B)_{syn}$; and $V1-B = 1.0709 + 0.966(V1-B)_{syn}$. In
the $U-B$ panel, the cross indicates the data points for  the stars HD
177724 and HD 210419, which were excluded from the calibration. As in
the previous two figures, circled points show the positions of the three
most rapid rotators in the sample:  HD~87901, HD~177724, and
HD~210419.
\label{figGENEVA}}
\end{figure}

%%%%%%%%%%%%%%%%%%%%%%%%%%%%%%% FIGURE 7 %%%%%%%%%%%%%%%%%%%%%%%%%%%%%%%%%%

\begin{figure}[ht]
\figurenum{7}
\epsscale{0.333}
\plotone{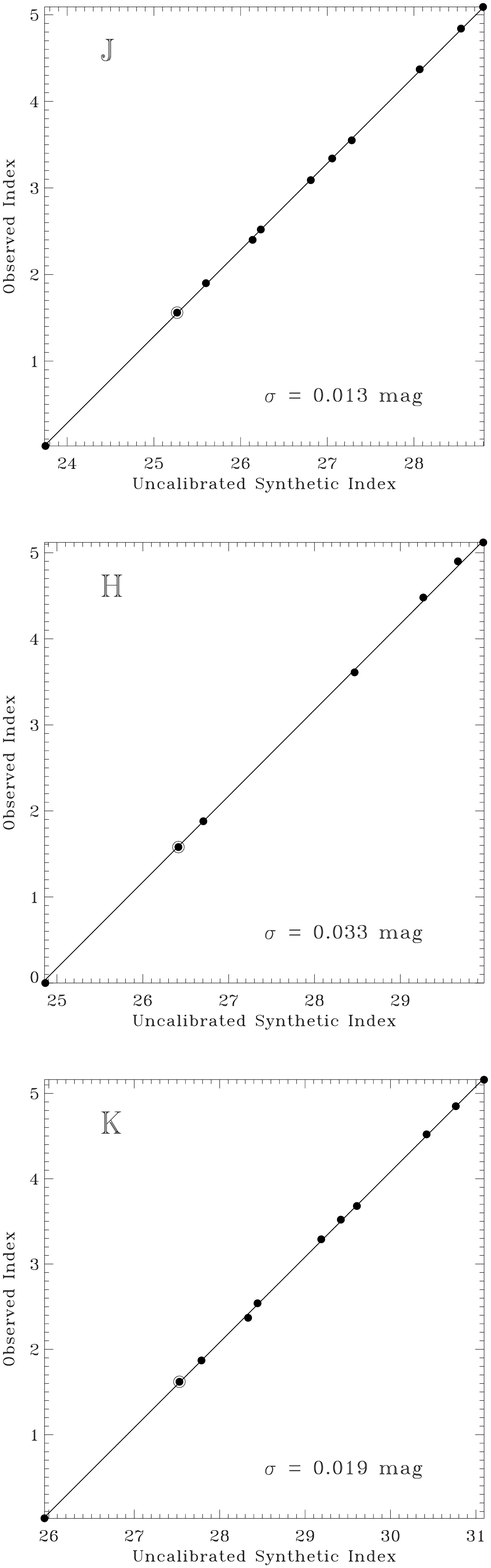}
\caption{Same as Figure \ref{figUBVRI}, but for the Johnson near-IR
magnitudes. The relationships between the observed and synthetic values
are as follows:  $J = -23.715 + J_{syn}$; $H = -24.827 + H_{syn}$; and
$K = -25.918 +K_{syn}$. As in the previous three figures, circled points
show the positions of the three most rapid rotators in the sample:
HD~87901, HD~177724, and HD~210419.
\label{figJHK}}
\end{figure}

%%%%%%%%%%%%%%%%%%%%%%%%%%%%%%% FIGURE 8 %%%%%%%%%%%%%%%%%%%%%%%%%%%%%%%%%%

\begin{figure}[ht]
\figurenum{8}
\epsscale{0.9}
\plotone{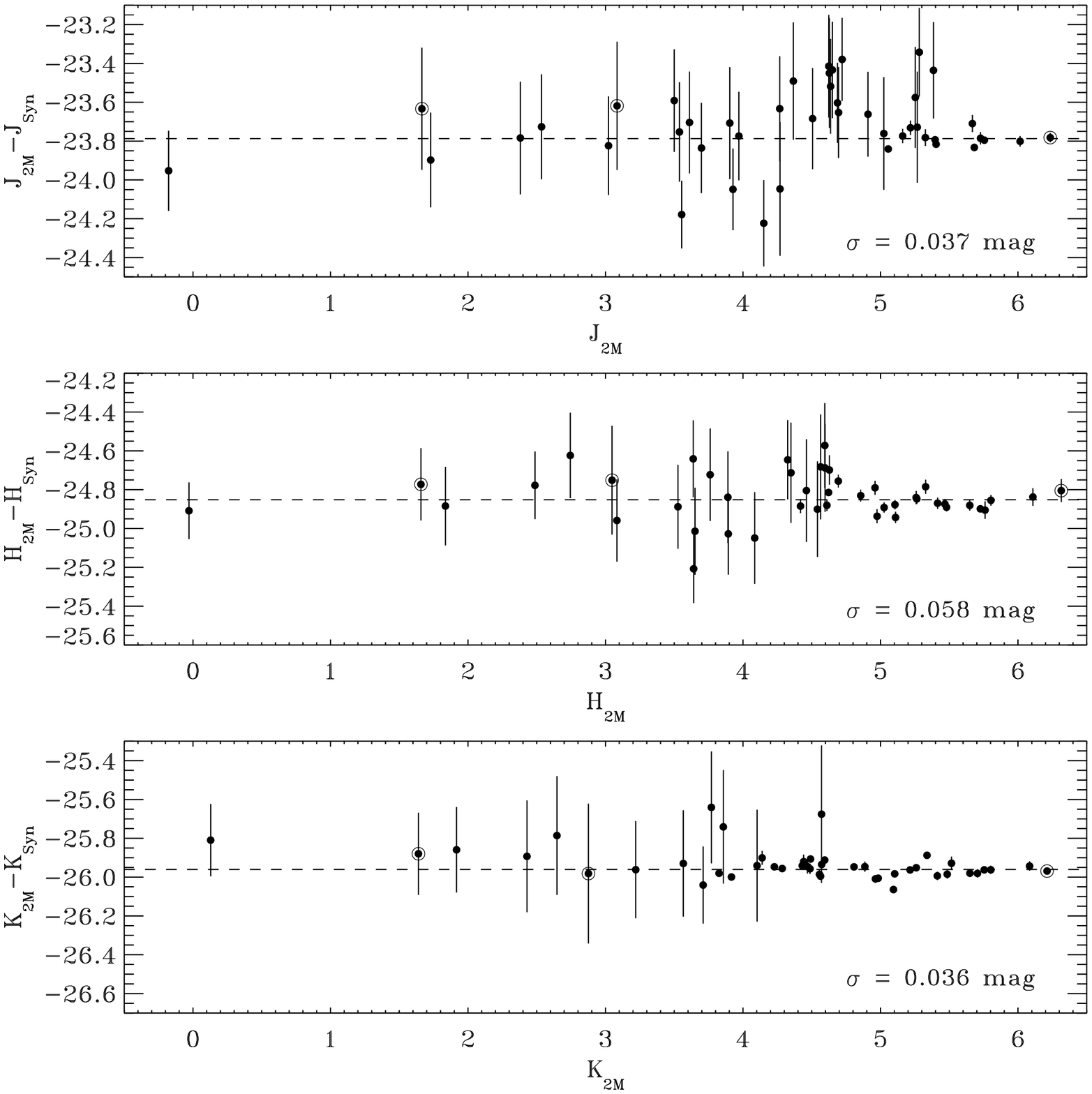}
\caption{Filter calibration relations for \tmass\/ photometry.  Each
panel shows the difference between the observed and synthetic
magnitudes, plotted against the observations.  Vertical error bars show
the individual observational errors, as listed in Table
\ref{tab2MASS}.  The dashed lines show the simple mean value of the
difference, based only on the observations with errors less than 0.1
mag.  The mean calibration relations are: $J_{2M} = -23.787
+J_{2Msyn}$; $H_{2M} = -24.852 + H_{2Msyn}$; and $K_{2M} = -25.961 +
K_{2Msyn}$.  The rms scatter about these relations is also shown in
each panel, based only on the data with errors less than 0.1 mag.
As in the previous four figures, circled points
show the positions of the three most rapid rotators in the sample:
HD~87901, HD~177724, and HD~210419.
\label{fig2MASS}}
\end{figure}

%%%%%%%%%%%%%%%%%%%%%%%%%%%%%%% FIGURE 9 %%%%%%%%%%%%%%%%%%%%%%%%%%%%%%%%%%

\begin{figure}[ht]
\figurenum{9}
\epsscale{0.80}
\plotone{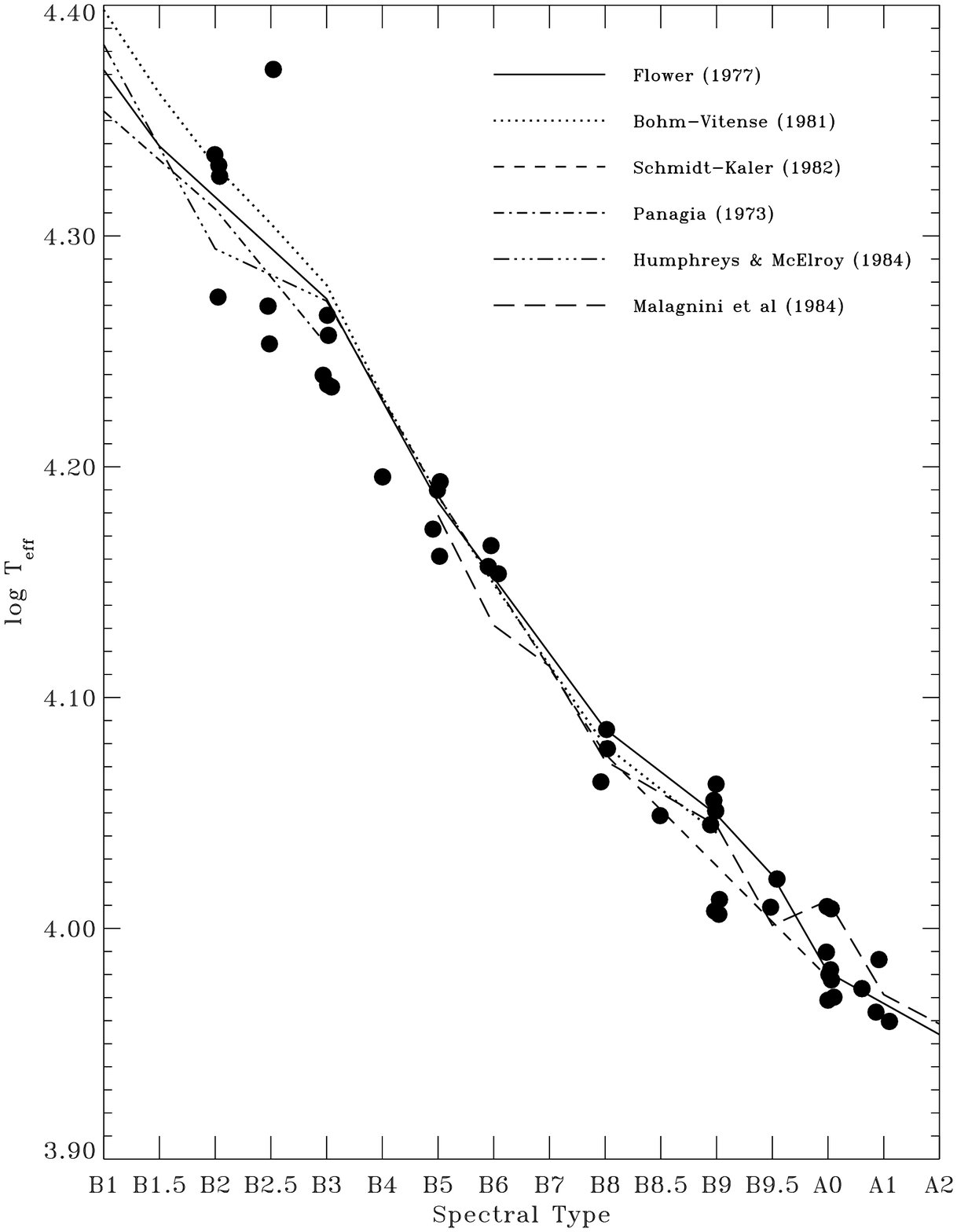}
\caption{ $\log T_{eff}$ as a function of spectral type for the program
stars (filled circles). The various dotted, dashed, and solid lines show a number of
published spectral type vs. $T_{eff}$ calibrations, as indicated in the
figure. Small random horizontal offsets have been added to the data points to increase their visibility \label{spty}}
\end{figure}

%%%%%%%%%%%%%%%%%%%%%%%%%%%%%%% FIGURE 10 %%%%%%%%%%%%%%%%%%%%%%%%%%%%%%%%%%

\begin{figure}[ht]
\figurenum{10}
\epsscale{1.1}
\plottwo{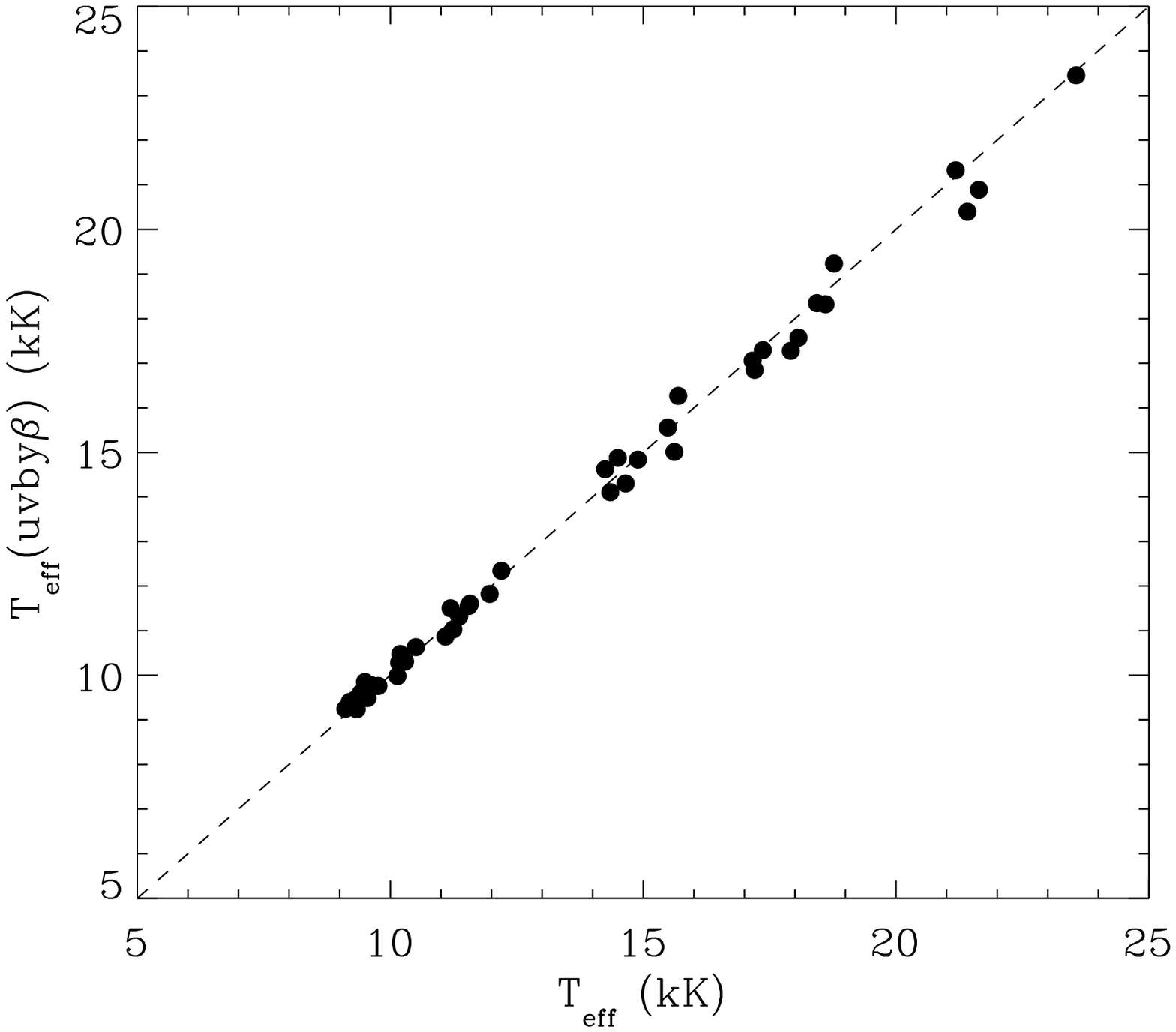}{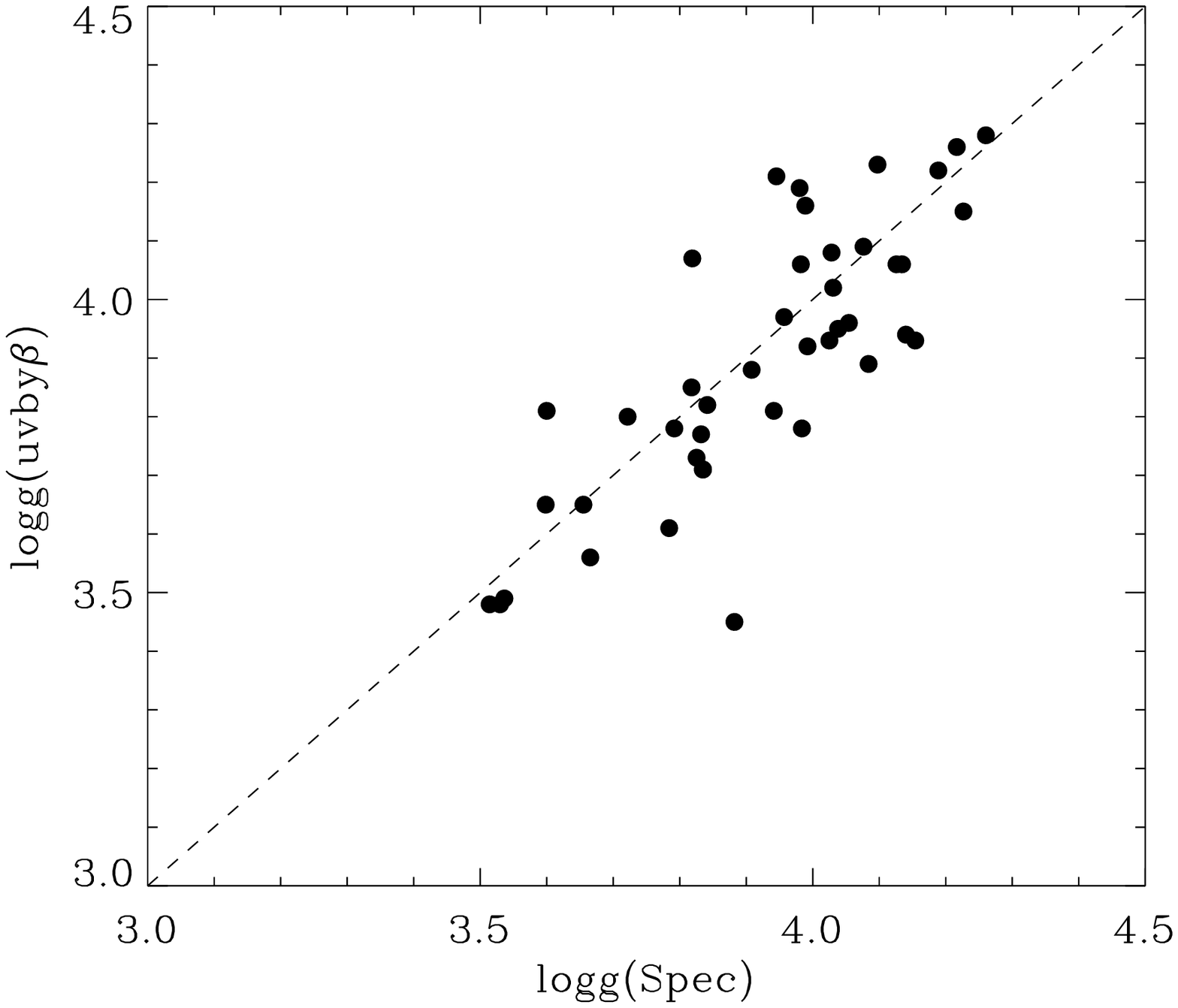}
\caption{Relations between the values of \teff\ and \logg\ derived in
this paper (see Table 5) plotted against those determined from the observed
uvby$\beta$ photometry and the calibration derived by Napiwotzki et
al.\ (1993).  The dashed line in each plot has a slope of unity,
indicating exact agreement.
\label{compare}}
\end{figure}

%%%%%%%%%%%%%%%%%%%%%%%%%%%%%%% FIGURE 11 %%%%%%%%%%%%%%%%%%%%%%%%%%%%%%%%%%

\begin{figure}[ht]
\figurenum{11}
\epsscale{0.80}
\plotone{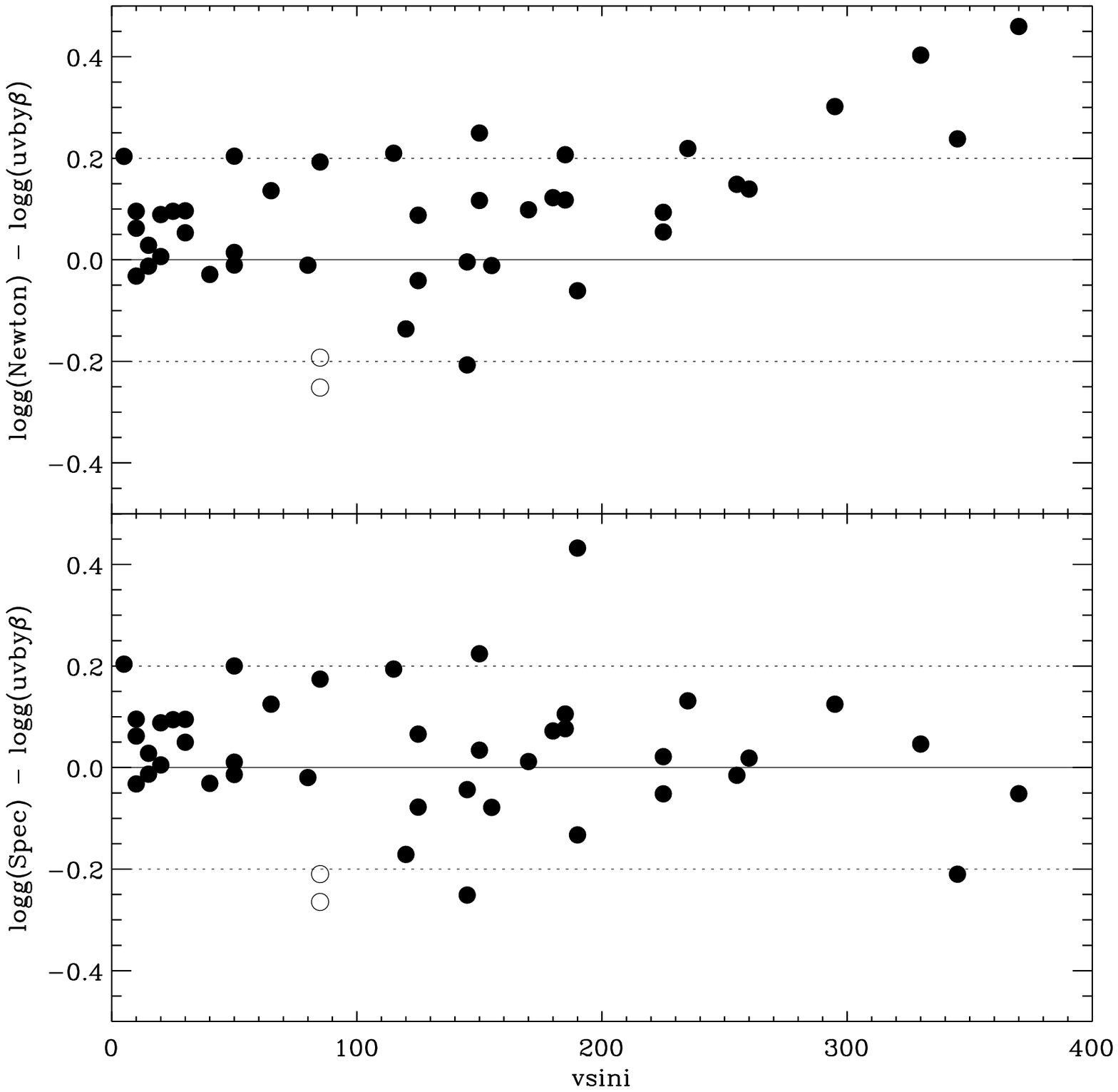}
\caption{Systematic effects in the \logg\/ determinations.  The upper
panel shows the \vsini\/ dependence of the differences between
Newtonian surface gravities determined from our \hip\/-constrained data
and the surface gravities based on $uvby\beta$ photometry.  The lower
panel shows how this systematic trend is removed by including
centrifugal effects in the computation of $\log g(Spec)$ (see Eq.
\ref{gspec}). Open symbols are shown for the stars HD 153808 and HD
199081, which are double-line spectroscopic binaries.  For such cases,
our \hip-based approach underestimates the Newtonian gravities by
approximately a factor-of-two, accounting for the large negative
residuals.  The star with the largest positive residual in the lower
plot, i.e., HD 225132 at $v \sin i = 190$ km/sec, lies outside the
bounds of the upper plot. The dotted lines in each panel at $\pm0.2$
are provided for visual reference.
\label{scatter}}
\end{figure}

\end{document}